\documentclass[12pt]{article}
\usepackage[dvips]{graphicx}
\usepackage{latexsym}
\usepackage{amssymb,amsfonts,amsmath}
\usepackage{graphicx}
\usepackage{indentfirst}
 \usepackage{bbm}
\usepackage{mathrsfs}

\usepackage{amsmath, amsthm,amssymb}
\usepackage{mathrsfs}
\usepackage{hyperref}
\usepackage{amsfonts}
\usepackage{dsfont}
\usepackage{slashed, tensor}
\usepackage{booktabs}
\usepackage{graphicx}
\usepackage{mathrsfs}

\parskip=\medskipamount

\usepackage{lipsum}
\usepackage{xargs}

\usepackage[colorinlistoftodos,prependcaption,textsize=tiny]{todonotes}
\newcommandx{\unsure}[2][1=]{\todo[linecolor=red,backgroundcolor=red!25,
bordercolor=red,#1]{#2}}
\newcommandx{\change}[2][1=]{\todo[linecolor=blue,
backgroundcolor=blue!25,bordercolor=blue,#1]{#2}}
\newcommandx{\STinfo}[1]{\todo[backgroundcolor=red!25,bordercolor=red,noline]{S.T.:#1}}
\newcommandx{\SKinfo}[1]{\todo[backgroundcolor=blue!25,bordercolor=blue,noline]{S.K.:#1}}

\renewcommand{\theequation}{\thesection.\arabic{equation}}
\newcommand{\be}{\begin{equation}}
\newcommand{\ee}{\end{equation}}
\newcommand{\bea}{\begin{eqnarray}}
\newcommand{\eea}{\end{eqnarray}}

\newcommand{\ba}{\begin{array}}
\newcommand{\ea}{\end{array}}

\def\double #1{#1{\hbox{\kern-2pt $#1$}}}

\newcommand{\bsubeq}{\begin{subequations}}
\newcommand{\esubeq}{\end{subequations}}

\numberwithin{equation}{section}

\begin{document}
\begin{titlepage}
\begin{flushright}
\end{flushright}
\vspace{5mm}

\begin{center}
{\Large \bf
Geometrical structure of Weyl invariants for spin three gauge field in general gravitational background in $d=4$
}
\\
\end{center}

\begin{center}

{\bf Ruben Manvelyan , Gabriel Poghosyan
} \\
\vspace{5mm}
\footnotesize{{\it Yerevan Physics Institute, Alikhanian Br. St. 2, 0036 Yerevan, Armenia}}
\vspace{2mm}
~\\

\end{center}

\begin{abstract}
\baselineskip=14pt
We construct all possible Weyl invariant actions in $d=4$ for linearized  spin three field in a general gravitational background.
The first action is  obtained as the square of the generalized Weyl tensor for a spin three gauge field in nonlinear
gravitational background. It is, however, not invariant under spin three gauge transformations.
We then construct two other nontrivial  Weyl but not gauge invariant actions which are linear in the Weyl tensor of the
background geometry.  We then discuss existence and uniqueness  of a possible linear combination of these three actions
which is gauge invariant. We do this at the linear order in the background curvature for Ricci flat backgrounds.
\end{abstract}

\vfill

\vfill
\end{titlepage}

\allowdisplaybreaks

\section{Introduction}
Conformal gravity has attracted considerable attention during the last thirty years \cite{FT}-\cite{Metsaev:2016rpa},
parallel to the development of higher spin gauge field theories \cite{Fronsdal1}, \cite{Vasiliev} \footnote{In this rather
technical contribution we do not pretend to cover all relevant references but only mention those which were
important for our understanding of the issues involved here}. It remained  an intriguing task to combine these two
developments and to construct an interacting higher spin conformal gauge field theory.
These two generalizations and extensions of ordinary gauge and gravity theories share many properties and
problems related to the high level and complicated structure of gauge symmetries \cite{Fradkin:1989yd}-\cite{Joung:2015jza}
and the necessity to include higher derivatives \cite{Kaparulin:2014vpa}-\cite{Joung:2016naf} which raises the
issue of the unitarity (see the discussions in \cite{Maldacena:2011mk}-\cite{Joung:2012qy}).
The interest in these intriguing topics intensified during the last decade with new
applications of conformal higher spin theories in the context of the  AdS/CFT correspondence.
Furthermore, the remarkable trivialization of the partition function in flat space
\cite{Joung:2015eny}-\cite{Tseytlin:2013jya} could be explained by the high level of gauge symmetry.
The possibility to obtain the exact partition function in some conformal higher spin field configurations
could prove useful for future nontrivial checks of the AdS/CFT conjecture.
Studying conformal higher spins is also helpful for the construction of couplings of higher spin gauge fields
to conformal currents \cite{Anselmi:1998bh}-\cite{Manvelyan:2004mb}.
This simple interaction allowed to perform some one loop
calculations and  to investigate the structure of conformal anomalies of higher spin fields \cite{Tseytlin:2013jya},
\cite{Manvelyan:2005ew}-\cite{Acevedo:2017vkk}.

In this paper we consider four-dimensional conformal higher spin (spin 3) theory in a general curved background.
We use the usual spin 2 Weyl symmetry (with scalar parameter) for the construction of generalized curvature and Weyl
tensors from the spin 3 field. We construct the linearized spin 3 Weyl tensor with three covariant derivatives and one
spin 3 field. The square of this tensor is our first conformal (Weyl) primary. This is done in a
general curved background, thus extending the usual flat space higher spin Weyl tensor of \cite{deWit} by additional terms
containing the background curvature.
Then, guided by ideas from spin 2 considerations and using the
technology developed in \cite{SUSI} related to the supersymmetric case, we construct all possible
primaries with conformal weight -4, which are linear in the background Weyl tensor.
For this purpose we have constructed and investigated the
conformal properties of the whole hierarchy of generalized spin 3 Christoffel symbols \cite{deWit} in a curved background.

\emph{The main result of our paper is the derivation of a second nontrivial Weyl invariant (\ref{final})}.
The existence of this additional primary, quadratic in the generalized spin 3 Christoffel symbols \cite{deWit} and linear in
the gravitational Weyl tensor, opens up the possibility to construct a unique Lagrangian which,  besides being
invariant under spin 2 and spin 3 Weyl transformations, is also invariant under spin-3 gauge transformations.
This will also involve the more trivial Weyl primary (\ref{3.21}) and the square of the spin 3 Weyl tensor.
Unfortunately, for the time being,  we were only able to prove invariance in Ricci (and therefore also Bach) flat backgrounds and to
linear order in the background Weyl tensor. Even to achieve this required computer assistance.
An interesting observation is that the combination obtained from the requirement of
spin 2 Weyl symmetry and spin 3 gauge symmetry is automatically invariant under the spin 3 Weyl symmetry,
where the spin 3 Weyl transformation is paramterized by a vector field and shifts of the trace of spin 3 gauge field.

More general backgrounds might be possible in the future. In this case one is confronted with the
results of \cite{Beccaria}\cite{Grigoriev} and also with discussions in \cite{Joung:2012qy} and later in  \cite{Nutma}.

In the next section we give a complete description of the well known spin 2 case: how to construct conformal primaries
and the construction of a unique gauge invariant action to all order on background curvature.
In Section 3 we apply this technology to the spin three case and obtain all necessary Weyl primaries with conformal
weight -4,  including the most nontrivial one mentioned above as our main results.
Then, in Section 4, we try to find a linear combination
of the three conformal primaries to obtain a gauge invariant Lagrangian.  This is a formidable task and we succeeded in
doing this only for
Ricci-flat backgrounds and to linear order in the background Weyl curvature.
Details of the construction are relegated to three appendices.

\section{Spin Two Example}

\quad In this section we demonstrate how the linearized conformal gravity action in $d=4$ can be obtained
from symmetry considerations alone. We show that possible Weyl invariant expressions  can be combined into a unique
gauge invariant action. To realize this idea we concentrate on the construction of possible primary fields
$\mathcal{L}_{\Delta}(g_{\alpha\beta},\nabla_{\lambda},h_{\mu\nu})$ with weight $\Delta$ with respect to local
Weyl transformations, written in terms of the linearized gravitational field $h_{\mu\nu}$ and covariant derivatives $\nabla_\mu$
in a general gravitational background $g_{\mu\nu}$:
\begin{equation*}
\delta \mathcal{L}_{\Delta}(g_{\alpha\beta},\nabla_{\lambda},h_{\mu\nu})
=\Delta \sigma(x)\mathcal{L}_{\Delta}(g_{\alpha\beta},\nabla_{\lambda},h_{\mu\nu}),
\end{equation*}
The Weyl transformations of background metric and the linearized spin two field are defined in a similar way
\begin{eqnarray*}
&&\delta h_{\mu\nu}(x)=2\sigma(x)h_{\mu\nu}(x),\\
&&\delta g_{\mu\nu}(x)=2\sigma(x)g_{\mu\nu}(x) .
\end{eqnarray*}
i.e. they are both symmetric spin two tensor primaries with conformal dimension $2$.

The most interesting primaries are scalars with conformal dimension $-4$, because they can be used to construct
Weyl invariant actions
\begin{equation}\label{00}
S^{\rm Weyl\,inv}=\int d^{4}x\sqrt{g}\mathcal{L}_{-4}(g_{\alpha\beta},\nabla_{\lambda},h_{\mu\nu}).
\end{equation}
Details of the notation and definitions can be found in Appendix A. Here we note only that it is convenient to introduce
the notation $\sigma_{\mu}=\partial_{\mu}\sigma(x)$ for the gradient of the scalar parameter of the Weyl symmetry.
Then e.g. formula (\ref{christ}) for the Weyl transformation of the Christoffel symbol can be written as
\begin{equation}\label{1.1}
\delta\Gamma^{\lambda}_{\mu\nu}= \sigma_{(\mu}\delta^{\lambda}_{\nu)}-g_{\mu\nu}\sigma^{\lambda}
\end{equation}
Using this we can investigate the Weyl transformation of second covariant derivatives of $h_{\mu\nu}$, where the four indices
have the symmetry of the Young Tableau "window" (curvature). For that we make symmetrization of indices of covariant derivatives $(\alpha\beta)$
and then perform antisymmetrization of two pairs of indices $[\alpha\mu]$ and $\{\beta\nu\}$\footnote{We use in this paper
$(\dots)$ and $<\dots>$ brackets for symmetrized sets and $[\dots]$ and $\{\dots\}$ for antisymetric pairs of indices.
No weight factor is included in the (anti)symmetrization, e.g. $v_{(\mu}w_{\nu)}=v_\mu w_\nu+v_\nu w_\mu$.}.
Then correcting this variation with terms proportional to background Schouten tensor  we  came to the following
nice transformation of the linearized curvature constructed from second derivatives:
\begin{eqnarray}
\mathcal{R}_{\alpha\mu,\beta\nu}&=&\frac{1}{4}[\{\nabla_{[\alpha},\nabla_{\{\beta}\}h_{\mu]\nu\}}-2K_{[\alpha\{\beta}h_{\mu]\nu\}}
-K^{\tau}_{[\alpha}g_{\mu][\beta}h_{\nu]\tau}-K^{\tau}_{[\beta}g_{\nu][\alpha}h_{\mu]\tau}],\label{1.6}\\
\delta \mathcal{R}_{\alpha\mu,\beta\nu}&=&2\sigma \mathcal{R}_{\alpha\mu,\beta\nu}
+\frac{1}{2}g_{[\alpha\{\beta} (\sigma^{\tau}\nabla_{\tau}h_{\mu]\nu\}}-\nabla_{(\mu]}h_{\nu\})\tau}\sigma^{\tau})\nonumber\\
&=&2\sigma \mathcal{R}_{\alpha\mu,\beta\nu}- g_{[\alpha\{\beta}\sigma^{\tau}\mathcal{G}_{\tau;\mu]\nu\} },\label{1.7}
\end{eqnarray}
where
\begin{eqnarray}
\mathcal{G}_{\tau;\mu\nu} &=&\frac{1}{2} (\nabla_{(\mu}h_{\nu)\tau}-\nabla_{\tau}h_{\mu\nu})\label{1.9}
\end{eqnarray}
is the linearized Christoffel symbol.
To obtain primary fields we investigate the Weyl transformations of the linearized Ricci tensor and Ricci scalar, obtained
as traces of (\ref{1.7}). Then we find that the following combination transforms as
\begin{eqnarray}
\delta \mathcal{K}_{\mu\nu}&=& \delta \frac{1}{2}(\mathcal{R}_{\mu\nu}-\frac{g_{\mu\nu}}{6}\mathcal{R})
=- \sigma^{\tau}\mathcal{G}_{\tau;\mu\nu}.\label{1.12}
\end{eqnarray}
This allows us to integrate (\ref{1.7}) in the form
\begin{eqnarray}
&& (\delta-2\sigma) \mathcal{R}_{\alpha\mu,\beta\nu}=g_{[\alpha\{\beta}\delta \mathcal{K}_{\mu]\nu\}}
=(\delta-2\sigma)(g_{[\alpha\{\beta}\mathcal{K}_{\mu]\nu\}})
\nonumber\\\label{1.13} \end{eqnarray}
We have thus constructed invariant linearized Weyl tensor (a primary field under Weyl transformations):
\begin{eqnarray}
\mathcal{W}_{\alpha\mu,\beta\nu} &=&\mathcal{R}_{\alpha\mu,\beta\nu}-g_{[\alpha\{\beta}\mathcal{K}_{\mu]\nu\}}.\label{1.14}
\end{eqnarray}
Note also that substraction of traces from (\ref{1.6}) or from the same expression but without last two terms with background
metric leads to the same results. So we can conclude that linearized Weyl tensor in general background is:
\begin{eqnarray}
&& \mathcal{W}_{\alpha\mu,\beta\nu}=\frac{1}{4}[\{\nabla_{[\alpha},\nabla_{\{\beta}\}h_{\mu]\nu\}}
-2K_{[\alpha\{\beta}h_{\mu]\nu\}}]-\textnormal{traces},
\nonumber\\\label{1.22}
\end{eqnarray}
and it is a conformal primary:
\begin{eqnarray}
&& \delta\mathcal{W}_{\alpha\mu,\beta\nu}=2 \sigma(x)\mathcal{W}_{\alpha\mu,\beta\nu}.
\end{eqnarray}
The background Weyl tensor is also $\Delta=2$ primary but without dependence on $h_{\mu\nu}$.
Having these two primaries we can construct:

1) One linear in linearized spin two field relevant ($\Delta=-4$)
primary
\begin{eqnarray}
\mathcal{L}^{lin} _{-4} &=& W^{\alpha\mu,\beta\nu}\mathcal{W}_{\alpha\mu,\beta\nu}=2 W^{\alpha\mu,\beta\nu}(\nabla_{\alpha}\nabla_{\beta}-K_{\alpha\beta})h_{\mu\nu},\label{1.23}
\end{eqnarray}
and corresponding invariant action produces correct equation of motion with Bach tensor for background metric:
\begin{eqnarray}
 B^{\mu\nu}=(\nabla_{\alpha}\nabla_{\beta}-K_{\alpha\beta})W^{\alpha\mu,\beta\nu}=0 . \label{1.24}
\end{eqnarray}

2) One four derivative quadratic primary
\begin{eqnarray}
\mathcal{L}^{\mathcal{W}^{2}}_{-4} &=&\frac{1}{2}\mathcal{W}^{\alpha\mu,\beta\nu}\mathcal{W}_{\alpha\mu,\beta\nu}.\label{1.21}
\end{eqnarray}

3) And several two and zero derivative primaries quadratic in linearized field

\begin{eqnarray}
\mathcal{L}^{\mathcal{W}W}_{-4} &=& \mathcal{W}^{\alpha\mu,\beta\nu}W_{\alpha\mu,\beta\nu}h^{\rho}_{\rho}, \label{ww1}\\
\mathcal{L}^{(1)W^{2}}_{-4} &=& W^{\alpha\mu,\beta\nu}
W_{\alpha\mu,\beta\nu}h_{\rho\tau}h^{\rho\tau}, \label{ww2}\\
\mathcal{L}^{(2)W^{2}}_{-4} &=& W_{\alpha\mu,\beta\nu}W_{\alpha\mu,\beta\nu}h^{\rho}_{\rho}h^{\tau}_{\tau}.\label{ww3}
\end{eqnarray}

We now turn to the Weyl variation of the linearized Christoffel symbol (\ref{1.9})
\begin{eqnarray}
(\delta-2\sigma)\mathcal{G}_{\tau;\mu\nu} &=&-\sigma_{\tau}h_{\mu\nu}+g_{\mu\nu}h_{\tau\lambda}\sigma^{\lambda}.\label{1.15}
\end{eqnarray}
We see that traceless in $\mu,\nu$ part of Christoffel symbol transforms in a way that quantity $\sigma_{\tau}$ arises only
with first $\tau$ index of symbol. Then taking in to account transformation low (\ref{1.12}) we can guess the last nontrivial
primary  with four derivatives and second order on spin three gauge field:
\begin{eqnarray}
\mathcal{L}^{W\mathcal{G}^{2}}_{-4}&=&\frac{1}{2}W^{\alpha\mu,\beta\nu}
\Big(\mathcal{G}_{\tau;\alpha\beta}\mathcal{G}^{\tau}_{\,\,\,;\mu\nu}
-2 h_{\alpha\beta}\mathcal{K}_{\mu\nu}\Big), \label{1.18}
\end{eqnarray}
with conformal weight -4:
\begin{equation}
\delta \mathcal{L}^{W\mathcal{G}^{2}}_{-4}=-4\sigma(x)\mathcal{L}^{W\mathcal{G}^{2}}_{-4}
\end{equation}

Now we consider the linearized gauge invariance:
\begin{eqnarray}
&& \bar{\delta} h_{\mu\nu}=\nabla_{(\mu}\epsilon_{\nu)}.\label{1.32}
\end{eqnarray}
The main goal now is to find unique  gauge invariant combination of the primaries presented above.
To find that we start from gauge variation of the last one and try to cancel at least some part from variation of (\ref{1.21}).
Immediately we see that cancelation can be observed only up to total derivative terms and therefore gauge invariance exists
only on the level of Weyl invariant actions (\ref{00}) where corresponding Lagrangians are our $-4$  weight primaries
(\ref{1.21})-(\ref{ww3}) and (\ref{1.18}). Doing in this direction and hiding long calculation in appendix B, we arrive to the
following unique "gauge invariant" combination of our primaries
\begin{equation}\label{GI}
\mathcal{L}^{GI}_{-4}=\mathcal{L}^{W\mathcal{G}^{2}}_{-4}
+\frac{1}{4}\mathcal{L}^{\mathcal{W}^{2}}_{-4}
- \frac{1}{16}\mathcal{L}^{\mathcal{W}W}_{-4}
+\frac{1}{32}\mathcal{L}^{(1)W^{2}}_{-4}
-\frac{1}{64}\mathcal{L}^{(2)W^{2}}_{-4}
\end{equation}
with the property that corresponding action transforms in respect to gauge transformation (\ref{1.32})
as follows\footnote{Here $\mathcal{L_{\epsilon}}$ is Lie derivative in direction of gauge vector parameter
$\epsilon^{\mu}$ in background metric $g_{\alpha\beta}$ defined by rule
$\mathcal{L_{\epsilon}} T^{\mu}_{\nu}=\epsilon^{\tau}\nabla_{\tau}T^{\mu}_{\nu}-\nabla_{\tau}\epsilon^{\mu}T^{\tau}_{\nu}
+\nabla_{\nu}\epsilon^{\tau}T^{\mu}_{\tau}$}
\begin{equation}\label{main}
\bar{\delta}\int d^{4}x\sqrt{g}\mathcal{L}^{GI}_{-4}
= \int d^{4}x\sqrt{g}\Big\{-\frac{1}{2}B^{\alpha\beta}\mathcal{L_{\epsilon}}h_{\alpha\beta}\Big\} .
\end{equation}
Therefore we prove that in the background with zero Bach tensor (conformal gravitational background)
gauge and Weyl invariant action is:
\begin{align}\label{1.64}
S_{GI}&= \frac{1}{8}\int d^{4}x\sqrt{g}\mathcal{W}^{\alpha\mu,\beta\nu}\mathcal{W}_{\alpha\mu,\beta\nu}
+\frac{1}{2}\int d^{4}x\sqrt{g}W^{\alpha\mu,\beta\nu}\Big(\mathcal{G}_{\tau;\alpha\beta}\mathcal{G}^{\tau}_{\,\,\,;\mu\nu}
-2 h_{\alpha\beta}\mathcal{K}_{\mu\nu}\Big)\nonumber\\
& - \frac{1}{16}\int d^{4}x\sqrt{g}\Big\{\mathcal{W}^{\alpha\mu,\beta\nu}W_{\alpha\mu,\beta\nu}h^{\rho}_{\rho}
-\frac{1}{2}W^{\alpha\mu,\beta\nu}
W_{\alpha\mu,\beta\nu}[h_{\rho\tau}h^{\rho\tau}-\frac{1}{4}h^{\rho}_{\rho}h^{\tau}_{\tau}]\Big\}
\end{align}

Of course this action can be obtained from expansion of the action for conformal gravity:
\begin{eqnarray}
S_{W(G)} &=&\frac{1}{2}\int d^{4}x\sqrt{G}W^{\alpha\mu,\beta\nu}(G)W_{\alpha\mu,\beta\nu}(G)\label{1.19}
\end{eqnarray}
up to second order on fluctuation $h_{\mu\nu}$ around general background $g_{\mu\nu}$:
\begin{eqnarray}
G_{\mu\nu} &=&g_{\mu\nu}+ h_{\mu\nu} .\label{1.20}
\end{eqnarray}

\section{Linearized Weyl Tensor and Other Primaries for Spin 3 Field}
Now we construct  Weyl tensor for spin 3 field in general gravitational background\footnote{All notations are
explained in the Appendix A}.
To do that first of all we should define gravitational Weyl transformation for spin 3 field
setting $4$ for conformal dimension of third rank symmetric tensor $h_{\mu\nu\lambda}$:
\begin{equation}\label{2.1}
\delta h_{\mu\nu\lambda}(x)=4\sigma(x) h_{\mu\nu\lambda}(x).
\end{equation}
Using the same as in previous section convenient vector notation $\sigma_{\mu}=\partial_{\mu}\sigma(x)$ for gradient of scalar
parameter of Weyl symmetry and investigating Weyl variation of the third  covariant derivative of $h_{\mu\nu\lambda}$
with two set of three symmetrized indices, we arrive to the following basic object
\begin{gather}
		H_{\alpha\beta\gamma,\mu\nu\lambda}=
		\frac{1}{6}\nabla_{(\alpha}\nabla_\beta\nabla_{\gamma)}h_{\mu\nu\lambda}
		-\frac{2}{3}\nabla_{(\alpha}K_{\beta\gamma)}h_{\mu\nu\lambda}
		+\frac{4}{3}K_{\alpha\beta}\nabla_{\lambda}h_{\mu\nu\gamma}
		+\frac{4}{3}K_{\beta\gamma}\nabla_{\mu}h_{\nu\lambda\alpha}
		+\frac{4}{3}K_{\gamma\alpha}\nabla_{\nu}h_{\lambda\mu\beta}\label{h3} .
	\end{gather}
Weyl transformation of which include \emph{only first derivative of scalar parameter} $\sigma_{\mu}$.
Another remarkable properties of this object is that after antisymmetrization over the index pairs $\{\alpha,\mu\}$,$\{\beta,\nu\}$
and $\{\gamma,\lambda\}$ we obtain index properties of Young Tableau with two row and three column and call it Riemann
curvature for spin three linearized gauge field:
\begin{eqnarray}
R_{\alpha\mu,\beta\nu,\gamma\lambda} &=& H_{\alpha\beta\gamma,\mu\nu\lambda}-H_{\mu\beta\gamma,\alpha\nu\lambda}
-H_{\alpha\nu\gamma,\mu\beta\lambda}+H_{\mu\nu\gamma,\alpha\beta\lambda}\nonumber\\
&-&H_{\alpha\beta\lambda,\mu\nu\gamma}+H_{\mu\beta\lambda,\alpha\nu\gamma}
+H_{\alpha\nu\lambda,\mu\beta\gamma}-H_{\mu\nu\lambda,\alpha\beta\gamma},\label{2.12}\\
R_{\alpha\mu,\beta\nu,\gamma\lambda}&=& R_{\beta\nu,\alpha\mu,\gamma\lambda}= R_{\alpha\mu,\gamma\lambda,\beta\nu}
= R_{\gamma\lambda,\beta\nu,\alpha\mu},\label{2.13}\\
R_{\alpha\mu,\beta[\nu,\gamma\lambda]}&=& R_{\alpha\mu,\beta\nu,\gamma\lambda}+ R_{\alpha\mu,\beta\gamma,\lambda\nu}
+ R_{\alpha\mu,\beta\lambda,\nu\gamma}=0 .\label{2.14}
\end{eqnarray}
The last condition is first Bianchi identity for spin 3 curvature.
Then we can observe \emph{that Weyl variation of curvature is linear on background metric}
\begin{eqnarray}
&&\delta R_{\alpha\mu,\beta\nu,\gamma\lambda}=
O(g;\sigma^{\tau};\nabla^{2};h).\label{2.15}
\end{eqnarray}
So we see that this, proportional to $g_{\alpha\beta}$, variation of $R_{\alpha\mu,\beta\nu,\gamma\lambda}$ can be completely
removed by subtraction of traces from curvature $R_{\alpha\mu,\beta\nu,\gamma\lambda}$. To subtract traces we note first that
due to symmetry properties we have unique first trace with one pair symmetric and another antisymmetric pair of indices and
satisfying  Bianchi identity obtaining from trace of (\ref{2.14})
\begin{eqnarray}
&& R_{\mu\nu,\gamma\lambda}=g^{\alpha\beta}R_{\alpha\mu,\beta\nu,\gamma\lambda},\label{2.16}\\
&& R_{\mu\nu,\gamma\lambda}=R_{\nu\mu,\gamma\lambda}=-R_{\mu\nu,\lambda\gamma},\label{2.17}\\
&&R_{\mu[\nu,\gamma\lambda]}=R_{\mu\nu,\gamma\lambda}+R_{\mu\gamma,\lambda\nu}+R_{\mu\lambda,\nu\gamma}=0.
\label{2.18}
\end{eqnarray}
The second trace we can take in two ways
\begin{eqnarray}
&& R^{(1)}_{\gamma\lambda}=g^{\mu\nu}R_{\mu\nu,\gamma\lambda}=-R^{(1)}_{\lambda\gamma}, \label{2.19}\\
&& R^{(2)}_{\nu\lambda}=g^{\mu\gamma}R_{\mu\nu,\gamma\lambda}=-R^{(2)}_{\nu\gamma} .\label{2.20}
\end{eqnarray}
Antisymmetrical properties of (\ref{2.20}) connected with the fact that it is second trace of (\ref{2.12}) with three antisymmetric
pairs of indices.
\begin{eqnarray}
&& R^{(2)}_{\nu\lambda}=g^{\alpha\beta}g^{\mu\gamma}R_{\alpha\mu,\beta\nu,\gamma\lambda} . \label{2.21}
\end{eqnarray}
Moreover the Bianchi identity (\ref{2.18}) relates  this two second traces:
\begin{eqnarray}
&& R^{(1)}_{\gamma\lambda}=R^{(2)}_{\gamma\lambda}-R^{(2)}_{\lambda\gamma}=2R^{(2)}_{\gamma\lambda} . \label{2.22}
\end{eqnarray}
To substruct traces from spin three curvatre we introduce analog of gravitational Schouten tensor in spin three case (d=4):
\begin{eqnarray}
K_{\mu\nu;\gamma\lambda} &=& \frac{1}{4}\Big[R_{\mu\nu;\gamma\lambda}-\frac{1}{10}\left(g_{\mu\nu}R^{(1)}_{\gamma\lambda}
+g_{[\gamma(\mu}R^{(2)}_{\nu)\lambda]}\right)\Big],\label{2.42}
\end{eqnarray}
with the same symmetry properties as (\ref{2.17}), (\ref{2.18})
and define  wanted spin three Weyl tensor in the form
\begin{eqnarray}
\mathcal{W}_{\alpha\mu,\beta\nu,\gamma\lambda} &=& R_{\alpha\mu,\beta\nu,\gamma\lambda}- g_{[\alpha\{\beta}K_{\mu]\nu\};
\gamma\lambda}- g_{[\alpha\{\gamma}K_{\mu]\lambda\};\beta\nu}-g_{[\gamma\{\beta}K_{\lambda]\nu\};\alpha\mu},\label{2.47}\\
\delta\mathcal{W}_{\alpha\mu,\beta\nu,\gamma\lambda} &=&4\sigma(x)\mathcal{W}_{\alpha\mu,\beta\nu,\gamma\lambda}.
\end{eqnarray}
In the full analogy with spin 2 case our spin 3 Weyl tensor is the same weight primary as the spin 3 gauge field but constructed from third covariant derivatives of it.
Squaring this sixth rank and $+4$ weigt primary tensor  and contracting with 6 metric tensors $g^{\mu\nu}$ of weght $-2$, we obtain first scalar primary Lagrangian with weight $-4$  relewant for Weyl invariant action:
\begin{equation}\label{3.1}
  L^{\mathcal{W}^{2}}_{-4}=\mathcal{\mathcal{W}}^{\alpha\mu,\beta\nu,\gamma\lambda}\mathcal{W}_{\alpha\mu,\beta\nu,\gamma\lambda}.
\end{equation}
And we know from spin 2 consideration that it is not enough for gauge invariance. This action is not gauge invariant in general
background: non invariance here is  proportional to the curvature of the background and arose from the commutators of derivatives
coming from definition of curvature with covariant derivative from gauge transformation of spin three field:
\begin{eqnarray}
&& \delta_{\epsilon} h_{\mu\nu\lambda}=\nabla_{\mu}\epsilon_{\nu\lambda} + \nabla_{\nu}\epsilon_{\lambda\mu}
  +\nabla_{\lambda}\epsilon_{\mu\nu}.
\end{eqnarray}
Therefore the expected solution of this problem is in the possible existence of other Weyl invariant primaries (but not gauge
invariant) with four or two covariant derivatives and first or second order on background curvature. Then the resulting combination
of Weyl invariant actions can compensate non invariance of (\ref{3.1}) supplemented with some reasonable restriction on
background metric $g_{\mu\nu}$. To make the next step in this direction, we start looking at Weyl noninvariant part of spin three curvature
written in terms of spin three Schouten tensor (\ref{2.42}) and understand  that the most interesting property of this tensor is it's
Weyl transformation:
\begin{eqnarray}
\delta K_{\mu\nu,\gamma\lambda}&=&2\sigma K_{\mu\nu,\gamma\lambda}
+ \frac{1}{3}\sigma^{\tau}\Gamma^{(2)}_{\tau[\gamma;\lambda]\mu\nu}
-\frac{1}{12} g_{[\gamma(\mu} \Gamma^{(2)\quad\,\alpha}_{\tau\alpha;\nu)\lambda]}\sigma^{\tau}\nonumber\\
&+&\frac{1}{24} g_{[\gamma(\mu}\Gamma^{(2)\quad\,\alpha}_{\tau\lambda];\nu)
\alpha}\sigma^{\tau}+\frac{1}{24} g_{[\gamma(\mu}\Gamma^{(2)\quad\,\alpha}_{\tau\nu);\lambda]\alpha}\sigma^{\tau}\label{3.17}
\end{eqnarray}
We see that this object transforms through the so called second spin three Christoffel symbol
$\Gamma^{(2)}_{\tau\gamma;\lambda\mu\nu}$ and it's traces.

So we should turn to the spin three analog of Christoffel symbols so called Freedman-deWitt hierarchy of Christoffel symbols for
higher spin case \cite{deWit}. The hierarchy means that in this spin three case we have in addition to spin three curvature two
other important objects: \emph{first and second Christoffel symbols with one and two covariant derivatives}.
First one is defined in unique way in general background:
\begin{eqnarray}
\Gamma^{(1)}_{\gamma;\mu\nu\lambda}&=&\nabla_{\gamma}h_{\mu\nu\lambda}-\nabla_{(\mu}h_{\nu\lambda)\gamma},\label{3.9}
\end{eqnarray}
with the following Weyl variation:
\begin{eqnarray}
\delta\Gamma^{(1)}_{\gamma;\mu\nu\lambda}&=&4\sigma \Gamma^{(1)}_{\gamma;\mu\nu\lambda}
+ 4 \sigma_{\gamma} h_{\mu\nu\lambda} -2 g_{(\mu\nu}h_{\lambda)\gamma\tau}\sigma^{\tau}.\label{3.10}
\end{eqnarray}
The second Christoffel symbol we can define through the first Christoffel symbol and curvature corrections in a way:
\begin{eqnarray}
&& \Gamma^{(2)}_{\beta\gamma;\mu\nu\lambda} = \nabla_{(\beta}\Gamma^{(1)}_{\gamma);\mu\nu\lambda}
- \frac{1}{2}\nabla_{(\mu}\Gamma^{(1)}_{<\beta;\gamma>\nu\lambda)}-8K_{\beta\gamma}h_{\mu\nu\lambda}
+2K_{(\mu<\beta}h_{\gamma>\nu\lambda)}\nonumber\\
&& +2g_{(\mu\nu}h_{\lambda)\tau <\beta}K^{\tau}_{\gamma>} -2g_{(\mu\nu}K^{\tau}_{\lambda)}h_{\tau \beta\gamma}
-g_{(<\beta(\mu}K^{\tau}_{\nu}h_{\lambda)\gamma>\tau}.\label{3.11}
\end{eqnarray}
In this case Weyl transformation of (\ref{3.11}) could be obtained only after enough
long but straightforward calculations So we arrive to the following transformation rule\footnote{Note that the last
term of (\ref{3.13}) can be also written through the first Christoffel symbol in some sophisticated way:
\begin{eqnarray}
6 g_{(\mu\nu}\nabla_{\lambda)}h_{\beta\gamma\tau}\sigma^{\tau}
&=& \frac{3}{2}g_{(\mu\nu}\sigma^{\tau}\Gamma^{(1)}_{[\lambda);\tau]\beta\gamma}
-\frac{3}{2}g_{(\mu\nu}\sigma^{\tau}\Gamma^{(1)}_{<\beta;\gamma>\lambda)\tau}\nonumber
\end{eqnarray}
but we prefer expression (\ref{3.13}) for simplicity.}
\begin{eqnarray}
\delta \Gamma^{(2)}_{\beta\gamma;\mu\nu\lambda}&=&4\sigma \Gamma^{(2)}_{\beta\gamma;\mu\nu\lambda}+  3 \sigma_{(\beta}
\Gamma^{(1)}_{\gamma);\mu\nu\lambda}\nonumber\\
&+&2 g_{\beta\gamma}\sigma^{\tau}\Gamma^{(1)}_{\tau;\mu\nu\lambda}
- g_{(\mu<\beta}\sigma^{\tau}\Gamma^{(1)}_{\tau;\gamma>\nu\lambda)}
+2 g_{(\mu\nu}\sigma^{\tau}\Gamma^{(1)}_{\tau;\lambda)\beta\gamma}\nonumber\\
&+& 6 g_{(\mu\nu}\nabla_{\lambda)}h_{\beta\gamma\tau}\sigma^{\tau}.\label{3.13}
\end{eqnarray}
Comparing (\ref{3.13}) with (\ref{1.15}) we see that our second Christoffel symbol (last in hierarchy) could play a role of usual
gravitational one and participate in construction of spin 3 generalization of the invariant  (\ref{1.18}).

In principal we have now all necessary ingredients to construct a spin three analog of invariant (\ref{1.18}). The obstacle to
construct it is only technical. But before we can easily construct analog of more simple primary (\ref{ww1}) where
linearized Weyl tensor contracted with background one and multiplied by trace of graviton field to get proper conformal weight.

It is clear that  to get scalar combination from six and four rank traceless tensors we need some antisymmetric second rank
tensor constructed from spin three field and one derivative. Looking at the traces of transformations of first Christoffel symbol
we can define the following primary:
\begin{eqnarray}
  L^{W\mathcal{W}}_{-4}&=&\frac{1}{4}\Gamma_{[\gamma,\lambda] \rho}^{(1)\quad \rho}
  \mathcal{W}^{\qquad\gamma\lambda}_{\alpha\mu,\beta\nu}W^{\alpha\mu,\beta\nu} - h^{\rho\lambda}_{\rho}\nabla^{\gamma}
  \mathcal{W}_{\gamma\lambda,\alpha\mu,\beta\nu}W^{\alpha\mu,\beta\nu},\nonumber\\\label{3.21}
\end{eqnarray}
where we used relation
\begin{eqnarray}
  \delta\nabla^{\gamma}\mathcal{W}_{\gamma\lambda,\alpha\mu,\beta\nu,} &=&2\sigma\nabla^{\gamma}\mathcal{W}_{\gamma\lambda,\alpha\mu,\beta\nu,}  +2\sigma^{\gamma}\mathcal{W}_{\gamma\lambda,\alpha\mu,\beta\nu}.\label{3.19}
\end{eqnarray}
To discover an analog of (\ref{1.18})  we should change to more irreducible objects in respect to Weyl transformation. These are traceless parts of our spin three field and Christoffel symbols defined above:
\begin{eqnarray}
  h^{T}_{\mu\nu\lambda}&=& h_{\mu\nu\lambda}-\frac{1}{6}g_{(\mu\nu}h^{\quad\alpha}_{\lambda)\alpha},\label{3.26}\\
  \Gamma^{(1)T}_{\gamma;\mu\nu\lambda} &=& \Gamma^{(1)}_{\gamma;\mu\nu\lambda}
  -\frac{1}{6}\Gamma^{(1)\alpha}_{\gamma;\alpha\,\,\,(\lambda}g_{\mu\nu)}, \label{3.24}\\
 \Gamma^{(1)T;T}_{[\gamma;\lambda]\mu\nu} &=& \Gamma^{(1);T}_{[\gamma;\lambda]\mu\nu}
 -\frac{1}{4}g_{[\gamma(\mu}\Gamma^{(1);T\,\,\,\alpha}_{\alpha;\nu)\lambda]},
\end{eqnarray}
and similar formulas for second Christoffel symbol:
 \begin{eqnarray}
 \Gamma^{(2)T}_{\beta\gamma;\lambda\mu\nu} &=& \Gamma^{(2)}_{\beta\gamma;\lambda\mu\nu}-\frac{1}{6}g_{(\mu\nu}\Gamma^{(2)\quad\alpha}_{\beta\gamma;\lambda)\alpha},\\
  \Gamma^{(2)T;T}_{\beta[\gamma;\lambda]\mu\nu} &=& \Gamma^{(2);T}_{\beta[\gamma;\lambda]\mu\nu}
 -\frac{1}{4}g_{[\gamma(\mu}\Gamma^{(2);T\quad\alpha}_{\beta\alpha;\nu)\lambda]},
 \end{eqnarray}
where $\Gamma_{(1)T;T}^{[\gamma;\lambda]\mu\nu}$ and $\Gamma_{(2)T;T}^{\beta[\gamma;\lambda]\mu\nu}$ are traceless in both pare of symmetric and anti symmetric indices\footnote{But index $\beta$ of second Christoffel is out of this game.}of (\ref{3.11}).
For full list of  Weyl transformations of traces of first two  Christoffel symbols and  corresponding traceless parts of them we refer to Appendix C. Here we present only the most important variations of traceless field and Christoffel symbols widely used for integration of new nontrivial invariant:
\begin{eqnarray}
&& \delta h^{T}_{\mu\nu\lambda}=4\sigma h^{T}_{\mu\nu\lambda},\label{3.579}\\
 && \delta\Gamma^{(1)T}_{\gamma;\mu\nu\lambda}=4\sigma\Gamma^{(1)T}_{\gamma;\mu\nu\lambda}+
  4\sigma_{\gamma}h^{T}_{\mu\nu\lambda},\label{3.580}\\ && \delta\Gamma^{(1)T;T}_{[\gamma;\lambda]\mu\nu}=4\sigma\Gamma^{(1)T;T}_{[\gamma;\lambda]\mu\nu}+
 4\sigma_{[\gamma}h^{T}_{\lambda]\mu\nu}- g_{[\gamma(\mu}h^{T}_{\nu)\lambda]\tau}\sigma^{\tau},\label{3.581}\\
 &&\delta\Gamma^{(2)T;T}_{\beta[\gamma;\lambda]\mu\nu}
 =4\sigma\Gamma^{(2)T;T}_{\beta[\gamma;\lambda]\mu\nu}+ 3\sigma_{\beta}\Gamma^{(1)T;T}_{[\gamma;\lambda]\mu\nu}\nonumber\\
 &&+3\Big(\sigma_{[\gamma}\Gamma^{(1)T}_{|\beta|;\lambda]\mu\nu}
 -\frac{1}{4}g_{[\gamma(\mu}\Gamma^{(1)T}_{|\beta|;\nu)\lambda]\tau}\sigma^{\tau}\Big)
+3\Big(g_{\beta[\gamma}\sigma^{\tau}\Gamma^{(1)T}_{\tau;\lambda]\mu\nu}
 -\frac{1}{4}g_{[\gamma(\mu}\sigma^{\tau}\Gamma^{(1)T}_{\tau;\nu)\lambda]\beta}\Big),\label{3.701}\\
 &&\delta\Gamma^{T}_{\mu\nu\lambda}= 2\sigma \Gamma^{T}_{\mu\nu\lambda} + 8\sigma^{\tau}\Gamma^{(1)T}_{\tau;\mu\nu\lambda},\label{3.681}
\end{eqnarray}
where we introduced notation \footnote{Note that $\Gamma^{T}_{\mu\nu\lambda}$ coincides in flat background with traceless part of Fronsdal equation and therefor it is gauge invariant in zero order on curvature.}
\begin{eqnarray}
\Gamma^{T}_{\mu\nu\lambda}&=&g^{\alpha\beta}\Gamma^{(2)T}_{\alpha\beta;\mu\nu\lambda}.\label{3.102}
\end{eqnarray}
Then subtracting trace from spin three Schouten tensor $K^{(T)}_{\mu\nu;\gamma\lambda}$ we see from (\ref{3.17}) that
\begin{equation}\label{3.95}
\delta K^{(T)}_{\mu\nu;\gamma\lambda}=\frac{1}{3}\sigma^{\beta}\Gamma^{(2)T;T}_{\beta[\gamma;\lambda]\mu\nu}.
\end{equation}
So we see that  traceless parts of some  field, Christoffel symbols and spin three Schouten tensor
\begin{equation}\label{set}
  h^{T}_{\mu\nu\lambda};\Gamma^{(1)T}_{\gamma;\mu\nu\lambda};\Gamma^{(1)T;T}_{[\gamma;\lambda]\mu\nu}; \Gamma^{(2)T;T}_{\beta[\gamma;\lambda]\mu\nu};\Gamma^{T}_{\mu\nu\lambda}; K^{(T)}_{\mu\nu;\gamma\lambda},
\end{equation}

transforms in close and more or less simple way in respect to Weyl transformations (\ref{3.579})-(\ref{3.681}) and we can use this property for construction of possible primaries. To illustrate this idea we should separate from (\ref{set}) a subset of two terms with close Weyl variations
$\{h^{T}_{\mu\nu\lambda};\Gamma^{(1)T}_{\gamma;\mu\nu\lambda}\}$ obtained from $h^{T}_{\mu\nu\lambda}$ and one covariant derivative. After that we construct primary combination by coupling it with similar  set formed from background Weyl and Cooton tensors $\{W^{\quad \gamma}_{\alpha\beta,\,\,\,\rho}; C_{\alpha\beta,\rho}\}$.
The most important point here is that the both sets start from primary field and end with nonprimary field constracted from previous one and from one covariant derivative. Then the cross coupled combination of them
\begin{equation}\label{3.28}
  W^{\quad \gamma}_{\alpha\beta,\,\,\,\rho}\Gamma^{(1)T}_{\gamma;\mu\nu\lambda}-C_{\alpha\beta,\rho}h^{T}_{\mu\nu\lambda}
\end{equation}
is the primary tensor with weight four. In the same spirit we can introduce another primary tensor with two set of three symmetrized indices:
\begin{gather}
 T^{\mu \nu \lambda}_{\alpha \beta \gamma}= \Gamma^{\tau,\mu \nu \lambda}_{(1)T}\Gamma^{(1)T}_{\tau,\alpha\beta\gamma}-\frac{1}{2}(h_{T}^{\mu \nu \lambda}\Gamma^{T}_{\alpha \beta \gamma}+h^{T}_{\alpha \beta \gamma}\Gamma^{\mu \nu \lambda}_{T}), \label{Ttensor}\\
 T^{\mu \nu }_{\alpha \beta }= T^{\mu \nu \lambda}_{\alpha \beta \lambda}, \quad\quad T^{\mu }_{\alpha}= T^{\mu \nu }_{\alpha \nu }.\label{Ttrace}
\end{gather}
This weight zero primary (\ref{Ttensor}) and corresponding first traces (\ref{Ttrace}) we incorporate with our whole set (\ref{set}) to construct the most nontrivial primary starting from background weyl tensor and square of second Christoffel tensor. This primary is playing role of analog of
(\ref{1.18}) in spin three case. The ideology of construction we described above and illustrated with some simple cases.\footnote{It is worth to note that contracting second trace of primary (\ref{Ttrace}) with background Bach tensor we quickly construct another Weyl invariant action:
\begin{equation}
 S^{W}_{B}=\frac{1}{2}\int d^{4}x\sqrt{g} B^{\alpha\beta}\Big[g^{\rho\tau} \Gamma^{(1)T}_{\rho;\alpha\mu\nu}\Gamma^{(1)T\mu\nu}_{\tau;\beta}-h^{T\mu\nu}_{\alpha} \Gamma^{T}_{\beta\mu\nu}\Big] \nonumber
\end{equation} }
But the proof is so long and complicated that we put it in Appendix C, presenting here only the \emph{final result for this primary}:

\begin{align}
L^{W\Gamma\Gamma}_{-4}
&=\frac{2}{3}W_{\mu \ ,\nu}^{\ \tau \ \, \rho}\tilde{\Gamma}^{(2)T,T}_{\beta[\gamma,\lambda]\tau \rho}\tilde{\Gamma}^{\beta[\gamma,\lambda]\mu\nu}_{(2)T,T}
+\frac{22}{9}W_{\gamma \ ,\mu}^{\ (\tau \ \, \rho)}\tilde{\Gamma}^{(2)T,T}_{\beta[\nu,\tau]\lambda \rho}\tilde{\Gamma}^{\beta[\gamma,\lambda]\mu\nu}_{(2)T,T}
-\frac{1}{6}W_{\gamma\lambda,}^{\ \ \tau \rho}\tilde{\Gamma}^{(2)T,T}_{\beta[\tau,\rho]\mu\nu}\tilde{\Gamma}^{\beta[\gamma,\lambda]\mu\nu}_{(2)T,T}\nonumber \\
&-\Big[\nabla_{\gamma}W_{\mu \ ,\nu}^{\ \tau \ \, \rho}-8\nabla_{\mu} W_{\nu\ ,\gamma}^{\ \tau \ \, \rho}+6C^{\ \rho}_{\mu\ \ ,\gamma}\delta^{\tau}_{\nu}\Big]\left( \frac{4}{3}\Gamma^{(1)T}_{\beta,  \lambda \tau \rho}\tilde{\Gamma}^{\beta[\gamma,\lambda]\mu\nu}_{(2)T,T}
-\frac{1}{2}\Gamma^{T}_{\lambda\tau\rho}\Gamma^{[\gamma,\lambda]\mu\nu}_{(1)T;T}-16h^{T}_{\lambda\tau\rho}K_{(T)}^{\mu\nu;\gamma\lambda}\right)\nonumber\\
& -\Big[12 W^{\quad\tau\quad\rho}_{\mu\quad,\nu}\Gamma^{(1)T,T}_{[\gamma;\lambda]\tau\rho}+44W^{\quad(\tau\quad\rho)}_{\gamma\quad,\mu}
\Gamma^{(1)T,T}_{[\nu;\tau]\lambda\rho} -3W^{\quad\tau\rho}_{\gamma\lambda,}\Gamma^{(1)T,T}_{[\tau;\rho]\mu\nu}\Big]
K_{(T)}^{\mu\nu;\gamma\lambda}\nonumber\\
& -2\Big[(\nabla^\sigma \nabla_\rho+4K^{\sigma}_{\rho})W^{\,\,\,\mu\,\,\,\,\,\nu}_{\alpha\,\, , \beta}\Big]T^{\alpha \beta \rho}_{\mu \nu \sigma} + \Big[4K^{\mu\tau}W_{\alpha\tau,\beta}^{\quad\,\,\nu}
 -3(\Box+2J)W_{\alpha\,\,\,, \beta}^{\,\,\,\mu\,\,\,\,\nu})\Big]T^{\alpha \beta}_{\mu \nu}, b\label{final}
\end{align}
where we introduced new notation:
\begin{equation}\label{tildegamma}
\tilde{\Gamma}^{\beta[\gamma,\lambda]\mu\nu}_{(2)T,T}=
\Gamma^{\beta[\gamma,\lambda]\mu\nu}_{(2)T,T}-\frac{3}{8}\left( g^{\beta[\gamma}\Gamma_{T}^{\lambda]\mu\nu}-\frac{1}{4}g^{[\gamma(\mu}\Gamma_{T}^{\nu)\lambda]\beta}\right),
\end{equation}
shifting second Christoffel symbol  by gauge invariant in zero order on background curvature terms. This modified
Christoffel symbol transforms without third line in (\ref{3.701}).

\section{ On Gauge Invariant Action for Conformal Spin Three}

In this section we address the final issue which is the  construction of a Weyl and gauge invariant action for a spin 3 field in
a background gravitational field. Following the prescription of \cite{SUSI} we can use approach
described here for spin two. In other words  we try to construct \emph{gauge invariant} combination of Weyl invariant actions
obtained from primary Lagrangians of the previous section. We should start from gauge variation
\begin{eqnarray}
  && \delta_{\epsilon} h_{\mu\nu\lambda}=\nabla_{\mu}\epsilon_{\nu\lambda} + \nabla_{\nu}\epsilon_{\lambda\mu}+\nabla_{\lambda}\epsilon_{\mu\nu}\label{4.1}
\end{eqnarray}
 of action
 \begin{equation}\label{W6W6}
   S_{\mathcal{W}^{2}}=\int d^{4}x\sqrt{g}L^{\mathcal{W}^{2}}_{-4},
 \end{equation}
 constructed from  square of spin three linearized Weyl tensor (\ref{3.1}) and try to cancel this  using partial integration with
 the variation of  action from second nontrivial invariant action (\ref{final})
\begin{equation}\label{W4G2G2}
 S_{W\Gamma\Gamma}=\int d^{4}x\sqrt{g}L^{\mathcal{W}^{2}}_{-4},
\end{equation}
integrating remaining terms to  other possible Weyl invariant part of action constructed from (\ref{3.21})
\begin{equation}\label{W4W6}
S_{\mathcal{W}W}=\int d^{4}x\sqrt{g}L^{\mathcal{W}W}_{-4}.
\end{equation}
In  contrast to Weyl invariance, this symmetry holds between Weyl primaries only up to total derivatives. We therefore use
actions instead of primary lagrangians.

The full check of gauge invariance is very tedious, and an analytical treatment to all orders in the background curvature is
out of reach. Instead with some experience we can guess the right combination and then check with the help of the computer.
For this we used the  Mathematica package xAct.
Following the spin two case, presented in the Appendix B, we expect that a particular linear combination of the three Weyl
invariant terms found in (\ref{W6W6}),
(\ref{W4G2G2}) and (\ref{W4W6}),  might be gauge invariant, at least to first order in background curvature.
We showed this, with the help of the computer, with the further assumption of a Ricci flat background. This implies that
Schouten and Bach tensors vanish and we therefore work to linear order in the Weyl tensor of the background geometry.
Even this turned out to be a formidable task with the result
\begin{gather}
\delta_{\epsilon}\Big[ S_{\mathcal{W}^{2}}-\frac{8}{5}S_{W\Gamma\Gamma}+\frac{4}{3}S_{W\mathcal{W}}\Big]=0+O(R^{2}, \, \,
K^{\alpha\beta},\, \, B)\label{4.2}
\end{gather}
The Lagrangians $S_{\mathcal{W}^{2}}$, $S_{W\Gamma\Gamma}$ and $S_{W\mathcal{W}}$ were given
in (\ref{3.1}), (\ref{final}) and (\ref{3.21}).
We choose this particular framework for two reasons: first it significantly shortens the computing time and second it
gives the possibility to express the terms in a way where the variation problem could  be reduced to the problem of
solving a system of linear equations.
Another long check of relation (\ref{4.2}) leads us to results that the parameter $\epsilon_{\mu\nu}$ in transformation (\ref{4.1})
could be non traceless as it should be in Fronsdal theory.
The last important relation is the following: The action we proposed in (\ref{4.2})
\begin{equation}\label{4.3}
 S_{CGI} = S_{\mathcal{W}^{2}}-\frac{8}{5}S_{W\Gamma\Gamma}+\frac{4}{3}S_{W\mathcal{W}}
\end{equation}
is not only conformal and gauge invariant in first order on background Weyl tensor, but invariant also in respect to spin 3 Weyl transformation (shifting of trace):
\begin{equation}\label{4.4}
 \delta_{\alpha} h_{\mu\nu\lambda}=g_{\mu\nu}\alpha_{\lambda} + g_{\nu\lambda}\alpha_{\mu}+g_{\lambda\mu}\alpha_{\nu}
\end{equation}
in the following way:
\begin{eqnarray}
   \delta_{\alpha}S_{\mathcal{W}^{2}}&=& 0 \label{4.5}\\
   \delta_{\alpha}\Big[S_{W\mathcal{W}}-\frac{6}{5}S_{W\Gamma\Gamma}\Big] &=& 0 \label{4.6}
\end{eqnarray}

\section{Conclusion and outlook}
In this paper we investigated the structure of Weyl covariant primaries in $d=4$. This primaries are relevant for using as a Weyl invariant Lagrangians,  expressed through the corresponding members of hierarchy of generalized Christoffel symbols \cite{deWit} and Weyl tensor for linearized spin 3 gauge field in general gravitational background. The main result is that in addition to the linearized spin 3 Weyl tensor corrected with background curvature terms we can construct additional nontrivial Weyl primary  in full analogy with spin 2 case. This primary is linear in background Weyl tensor and quadratic in linearized second Christoffel symbol. This Christoffel symbol is the last before curvature in corresponding hierarchy for spin 3 case \cite{deWit}. A possible combination of these primaries, in principal, can be interpreted as a gauge and Weyl invariant action with corresponding restriction on background geometry. This could be investigated in the future. Here we only briefly discuss  the possible combination of these invariants in linear on background Weyl tensor approximation using computer calculations.  Also it is reasonable in the future obtain connections with the results of \cite{Beccaria} where authors using another methods claim that for gauge invariance in the second order on background curvature we should introduce interaction with additional spin one field. Unfortunately by technical reason, at the moment, we can make some checking of gauge invariance using computer only in first order on background Weyl tensor. Another reasonable  task here to obtain better understanding of connections with the results of \cite{Nutma} and \cite{Grigoriev}.

\textbf{Acknowledgements:}

RM is grateful to Stefan Theisen and Sergei Kuzenko for many valuable discussions, useful conversations and support. RM also acknowledges hospitality of Albert Einstein Institute in Potsdam-Golm  when this project was initiated. The work of RM was supported in part by the Alexander von Humboldt Foundation and by the Science Committee of the Ministry of Science and Education of the Republic of Armenia under contract 15T-1C233.

\section{Appendix A. Notations and Conventions}
\setcounter{equation}{0}
\renewcommand{\theequation}{A.\arabic{equation}}
We work in a $d=4$ dimensional curved space with general metric $g_{\mu\nu}$ and use the following
conventions for covariant derivatives and curvatures:
\begin{eqnarray}
  \nabla_{\mu}V^{\rho}_{\lambda}&=& \partial_{\mu}V^{\rho}_{\lambda}+
  \Gamma^{\rho}_{\mu\sigma}V^{\sigma}_{\lambda}-\Gamma^{\sigma}_{\mu\lambda}V^{\rho}_{\sigma} , \\
  \Gamma^{\rho}_{\mu\nu} &=& \frac{1}{2} g^{\rho\lambda}\left(\partial_{\mu}g_{\nu\lambda}+
  \partial_{\nu}g_{\mu\lambda} - \partial_{\lambda}g_{\mu\nu}\right) , \\
  \left[\nabla_{\mu} , \nabla_{\nu}\right]V^{\rho}_{\lambda} &=&
  R^{\quad\,\,\rho}_{\mu\nu, \sigma}V^{\sigma}_{\lambda}
  -R^{\quad\,\,\sigma}_{\mu\nu, \lambda}V^{\rho}_{\sigma} ,\\
  R^{\quad\,\,\rho}_{\mu\nu, \lambda}&=& \partial_{\mu}\Gamma^{\rho}_{\nu\lambda}
  -\partial_{\nu}\Gamma^{\rho}_{\mu\lambda}+\Gamma^{\rho}_{\mu\sigma}\Gamma^{\sigma}_{\nu\lambda}
  -\Gamma^{\rho}_{\nu\sigma}\Gamma^{\sigma}_{\mu\lambda} ,\\
  R_{\mu\lambda}&=& R^{\quad\,\,\rho}_{\mu\rho\lambda}\quad , \quad
  R=R^{\,\,\mu}_{\mu} .
\end{eqnarray}
The corresponding local conformal transformations (Weyl rescalings)
\begin{eqnarray}
  \delta g_{\mu\nu}&=&2\sigma(x) g_{\mu\nu} , \quad\
  \delta g^{\mu\nu} = -2\sigma(x) g^{\mu\nu} ,\\
   \delta\Gamma^{\lambda}_{\mu\nu}&=& \partial_{\mu}\sigma\delta^{\lambda}_{\nu}
  +\partial_{\nu}\sigma\delta^{\lambda}_{\mu}-
  g_{\mu\nu}\partial^{\lambda}\sigma , \label{christ}\\
   \delta R^{\quad\,\,\rho}_{\mu\nu, \lambda}&=&\nabla_{\mu}
  \partial_{\lambda}\sigma\delta^{\rho}_{\nu}-
  \nabla_{\nu}\partial_{\lambda}\sigma\delta^{\rho}_{\mu}+
  g_{\mu\lambda}\nabla_{\nu}\partial^{\rho}\sigma
  - g_{\nu\lambda}\nabla_{\mu}\partial^{\rho}\sigma ,\\
  \delta R_{\mu\lambda}&=&(d-2)\nabla_{\mu}\partial_{\lambda}\sigma +
g_{\mu\lambda}\Box
\sigma ,\\
 \delta R&=& -2\sigma R + 2(d-1)\Box \sigma\,,
\end{eqnarray}
are first oder in the infinitesimal local scaling parameter
$\sigma$.

We then introduce the Weyl ($W$) and Schouten ($K$) tensors, as well as
the scalar $J$
\begin{eqnarray}
  R_{\mu \nu } &=& (d - 2)K_{\mu \nu }  + g_{\mu \nu}J , \quad J=\frac{1}{2(d - 1)}R\,\,, \label{A11}\\
  W_{\mu \nu, \lambda }^{\quad\,\,\rho}   &=& R_{\mu \nu, \lambda }^{\quad\,\,\rho}   - K_{\mu \lambda }
 \delta _\nu  ^\rho   +
 K_{\nu \lambda } \delta _\mu  ^\rho   - K_\nu  ^\rho  g_{\mu \lambda }  +
 K_\mu  ^\rho  g_{\nu \lambda } \,\,, \label{A12}\\
  \delta K_{\mu \nu } &=& \nabla _\mu  \partial _\nu
 \sigma \,,\,\,\,\, \label{A13}\\ \delta J&=&-2\sigma J + \Box
 \sigma\,,\,\,\,\,\,\label{deltaJ}\\
 \delta W_{\mu \nu, \lambda }^{\quad\,\,\rho}&=& 0\,,
\end{eqnarray}
which are more convenient because their conformal transformations
are "diagonal".

To describe the Bianchi identity with these tensors, we have to
introduce the so called Cotton tensor
\begin{eqnarray}
  C_{\mu \nu, \lambda } &=& \nabla _\mu  K_{\nu \lambda }  - \nabla _\nu
 K_{\mu \lambda } \,\,,
  \label{bi1}\\ \delta C_{\mu \nu, \lambda }
 &=&  - \partial _\alpha  \sigma W_{\mu \nu, \lambda
 }^{\quad\,\,\alpha}\,\,\,\,,\,\,\,\,\,  C_{[\mu \nu, \lambda ] = 0\,\,.}\label{A17}
\end{eqnarray}
All important properties of these tensors following from the Bianchi
identity can then be listed as
\begin{eqnarray}
  \nabla _{[\alpha } W_{\mu \nu ],\lambda }^{\quad\,\,\,\rho} &=& g_{\lambda [\alpha }
  C_{\mu \nu ],}^{\quad\,\,\rho}
   - \delta _{[\alpha }^\rho  C_{\mu \nu ],\lambda } \,\,,\, \label{A.18}\\
 \nabla _\alpha  W_{\mu \nu ,\lambda }^{\quad\,\,\alpha}&=& \left( {3
- d}\right)C_{\mu \nu ,\lambda } \,\,\,,\\
 \nabla^{\mu}K_{\mu\nu}&=&\partial_{\nu}J\,,\label{bi2}\\
 C_{\mu \nu, }^{\quad\,\nu} &=& 0\,\,,\quad\quad
\nabla^{\lambda}C_{\mu\nu,\lambda}=0\,.
\end{eqnarray}
Finally we introduce the last important conformal
object in the above listed hierarchy, namely the symmetric and traceless
Bach tensor
\begin{eqnarray}
B_{\mu\nu}&=&\nabla^{\lambda}C_{\lambda\mu, \nu}
  +K^{\lambda}_{\alpha}W_{\lambda\mu, \nu  }^{\quad\,\,\alpha} ,
\end{eqnarray}
whose conformal transformation and divergence are expressed in terms
of the Cotton and the  Schouten tensors as follows
\begin{eqnarray}
    \delta B_{\mu\nu}&=&-2\sigma B_{\mu\nu}+(d-4)\nabla^{\lambda}\sigma
  \left( C_{\lambda\mu,\nu}+C_{\lambda\nu,\mu}\right) ,\label{bachtr}\\
   \nabla^{\mu}B_{\mu\nu}&=&(d-4)C_{\alpha\nu,\beta}K^{\alpha\beta} .
  \end{eqnarray}
Note that only in four dimensions the Bach tensor is
conformal invariant and divergenceless.
We use also other dimension  dependent (d=4) relation for any traceless tensor with symmetries of Weyl tensor:
\begin{equation}\label{1.54}
  W^{[\alpha\mu}_{[\beta\nu}\delta^{\lambda]}_{\rho]}=0 .
\end{equation}
From last relation we can derive the following important identities:
\begin{eqnarray}
  W^{\alpha\mu,\beta\rho}W_{\alpha\mu,\beta\lambda} &=& \frac{1}{4}\delta^{\rho}_{\lambda}W^{\alpha\mu,\beta\nu}W_{\alpha\mu,\beta\nu}, \label{1.55}\\
  \mathcal{W}^{\alpha\mu,\beta(\rho}W_{\alpha\mu,\beta}^{\quad\quad\lambda)} &=& \frac{1}{2}g^{\rho\lambda}\mathcal{W}^{\alpha\mu,\beta\nu}W_{\alpha\mu,\beta\nu}, \label{1.56}\\
  \frac{1}{4}W^{\alpha\mu,\beta\nu}W_{\alpha\mu,\beta\nu}(h^{\rho}_{\rho}h^{\tau}_{\tau}-h_{\rho\tau}h^{\rho\tau}) &=& 2 h^{\rho\mu}h^{\lambda\nu}W^{\alpha\,\,\,\beta}_{\,\,\,\,\,\rho,\,\,\,\lambda}W_{\alpha\mu,\beta\nu}+ h^{\rho\beta}h^{\lambda\nu}W^{\alpha\mu}_{\,\,\,\,,\rho\lambda}W_{\alpha\mu,\beta\nu}\nonumber\\
  &-&2W^{\alpha\rho,\beta\lambda}h_{\rho\lambda}W_{\alpha\mu,\beta\nu}h^{\mu\nu}.\label{1.57}
\end{eqnarray}
\section{Appendix B. Spin Two Details}
\setcounter{equation}{0}
\renewcommand{\theequation}{B.\arabic{equation}}

To prove (\ref{main}) we start from gauge variation (\ref{1.32}) of the initial object
\begin{eqnarray}
  &&\textbf{R}_{\alpha\mu,\beta\nu}=\frac{1}{4}\{\nabla_{[\alpha},\nabla_{\{\beta}\}h_{\mu]\nu\}}.\label{1.33}
\end{eqnarray}
This can be rewritten in the following form
\begin{eqnarray}
  && \bar{\delta}\textbf{R}_{\alpha\mu,\beta\nu}= -\frac{1}{4}[R_{\alpha\mu,[\beta}^{\qquad\tau}\bar{\delta} h_{\nu]\tau} + R_{\beta\nu,[\alpha}^{\qquad\tau}\bar{\delta} h_{\mu]\tau}] + \mathcal{L_{\epsilon}}R_{\alpha\mu,\beta\nu},\label{1.34}
\end{eqnarray}
where $R_{\alpha\mu,\beta\nu}$ is background curvature and $\mathcal{L_{\epsilon}}$ is Lie derivative in direction of gauge parameter vector $\epsilon^{\tau}$:
\begin{eqnarray}
  &&\mathcal{L_{\epsilon}}R_{\alpha\mu,\beta\nu}=\epsilon^{\tau}\nabla_{\tau}R_{\alpha\mu,\beta\nu}
  +R_{\alpha\mu,\tau[\nu}\nabla_{\beta]}\epsilon^{\tau}+R_{\beta\nu,\tau[\mu}\nabla_{\alpha]}\epsilon^{\tau}.\label{1.35}
\end{eqnarray}
So we see that the following improved linearized curvature
\begin{eqnarray}
  && \mathcal{\bar{R}}_{\alpha\mu,\beta\nu}=\frac{1}{4}[\{\nabla_{[\alpha},\nabla_{\{\beta}\}h_{\mu]\nu\}}+R_{\alpha\mu,[\beta}^{\qquad\tau} h_{\nu]\tau} + R_{\beta\nu,[\alpha}^{\qquad\tau}h_{\mu]\tau}]\label{1.36}
\end{eqnarray}
transforms covariantly:
\begin{eqnarray}
  && \bar{\delta}\mathcal{\bar{R}}_{\alpha\mu,\beta\nu}=\mathcal{L_{\epsilon}}R_{\alpha\mu,\beta\nu}.\label{1.37}
\end{eqnarray}
Then after some calculations we can see that (\ref{1.36}) can be rewritten using linearized  Christoffel symbols (\ref{1.9}) in the form:
\begin{eqnarray}
  && \mathcal{\bar{R}}_{\alpha\mu,\beta\nu}=\nabla_{[\alpha}\mathcal{G}^{\tau}_{\mu]\beta}g_{\nu\tau}+R^{\quad\,\,\, \tau}_{\alpha\mu;\beta}h_{\nu\tau}\label{1.38}
\end{eqnarray}
and coincides with the linearized expansion of usual nonlinear curvature for general metric (\ref{1.20}).

Expanding background curvatures on Weyl and Schouten tensors we obtain the following relations:
\begin{eqnarray}
 \mathcal{\bar{R}}_{\alpha\mu,\beta\nu}&=&\mathcal{R}_{\alpha\mu,\beta\nu}+h_{[\alpha\{\beta}K_{\mu]\nu\}}+\frac{1}{4}[W_{\alpha\mu,[\beta}^{\qquad\tau} h_{\nu]\tau} + W_{\beta\nu,[\alpha}^{\qquad\tau}h_{\mu]\tau}],\label{1.39}\\
 \bar{\delta}\mathcal{R}_{\alpha\mu,\beta\nu}&=&\mathcal{L_{\epsilon}}W_{\alpha\mu,\beta\nu}+ g_{[\alpha\{\beta}\mathcal{L_{\epsilon}}K_{\mu]\nu\}}-\frac{1}{4}[W_{\alpha\mu,[\beta}^{\qquad\tau}\bar{\delta} h_{\nu]\tau} + W_{\beta\nu,[\alpha}^{\qquad\tau}\bar{\delta} h_{\mu]\tau}],\quad\quad\label{1.40}
\end{eqnarray}
where $\mathcal{R}$ here is the same as in (\ref{1.6})
\begin{equation}\label{1.41}
 \mathcal{R}_{\alpha\mu,\beta\nu}=\frac{1}{4}[\{\nabla_{[\alpha},\nabla_{\{\beta}\}h_{\mu]\nu\}}-2K_{[\alpha\{\beta}h_{\mu]\nu\}}
  -K^{\tau}_{[\alpha}g_{\mu][\beta}h_{\nu]\tau}-K^{\tau}_{[\beta}g_{\nu][\alpha}h_{\mu]\tau}].\\
\end{equation}
Then taking traces from (\ref{1.40}) we arrive to the following variations
\begin{eqnarray}
  \bar{\delta}\mathcal{R}_{\mu\nu}&=&g^{\alpha\beta}\bar{\delta}\mathcal{R}_{\alpha\mu,\beta\nu}=
  2\mathcal{L_{\epsilon}}K_{\mu\nu}-\frac{1}{2}W_{\alpha\mu,\beta\nu}\delta h^{\alpha\beta}+ g_{\mu\nu}(\delta h^{\alpha\beta}K_{\alpha\beta} + \mathcal{L_{\epsilon}}J),\quad \label{1.42}\\
  \bar{\delta}\mathcal{R}&=&g^{\mu\nu}\bar{\delta}\mathcal{R}_{\mu\nu}=6\bar{\delta}\mathcal{J}=6 (\delta h^{\alpha\beta}K_{\alpha\beta} + \mathcal{L_{\epsilon}}J),\label{1.43}\\
  \bar{\delta}\mathcal{K}_{\mu\nu}&=&\frac{1}{2}\bar{\delta}(\mathcal{R}_{\mu\nu}-g_{\mu\nu}\mathcal{J})=\mathcal{L_{\epsilon}}K_{\mu\nu}-\frac{1}{4}W_{\alpha\mu,\beta\nu}\delta h^{\alpha\beta},\label{1.44}
  \end{eqnarray}
  we obtain for linearized Weyl tensor (\ref{1.14})
\begin{eqnarray}
 \bar{\delta} \mathcal{W}_{\alpha\mu,\beta\nu} &=&\mathcal{L_{\epsilon}}W_{\alpha\mu,\beta\nu}-\frac{1}{4}[W_{\alpha\mu,[\beta}^{\qquad\tau} \bar{\delta} h_{\nu]\tau} + W_{\beta\nu,[\alpha}^{\qquad\tau}\bar{\delta} h_{\mu]\tau}
  - g_{[\alpha\{\beta}W_{\mu],\,\,\nu\}}^{\,\,\,\rho\quad\tau} \bar{\delta} h_{\rho\tau}].\quad\qquad\label{1.45}
\end{eqnarray}
Comparing (\ref{1.6}), (\ref{1.14}) and (\ref{1.38}), (\ref{1.39}) we see that linearized curvature obtained from gauge consideration connected with Weyl invariant traceless linearized Weyl tensor in the following way:
\begin{eqnarray}
  \mathcal{\bar{R}}_{\alpha\mu,\beta\nu} &=& \mathcal{W}_{\alpha\mu,\beta\nu} + g_{[\alpha\{\beta}\mathcal{K}_{\mu]\nu\}}+h_{[\alpha\{\beta}K_{\mu]\nu\}}+\frac{1}{4}[W_{\alpha\mu,[\beta}^{\qquad\tau} h_{\nu]\tau} + W_{\beta\nu,[\alpha}^{\qquad\tau}h_{\mu]\tau}].\nonumber\\\label{1.46}
\end{eqnarray}
This linearized curvature cares  all symmetry properties of nonlinear Riemann curvature and satisfies also to usual first Bianchi identity. The second (differential) Bianchi identity for linearized curvature looks like
\begin{align}
\nabla_{[\tau}\mathcal{\bar{R}}_{\alpha\mu],\beta\nu}
-\mathcal{G}^{\rho}_{\beta[\tau}R_{\alpha\mu],\rho\nu}-\mathcal{G}^{\rho}_{\nu[\tau}R_{\alpha\mu],\beta\rho}=0 . \label{1.47}
\end{align}
Using this and partial integration we can derive another important relation
\begin{align}
  \frac{1}{2}\int d^{4}x\sqrt{g}W^{\alpha\mu,\beta\nu} \mathcal{\bar{R}}_{\alpha\mu,\beta\tau}\nabla_{\nu}\epsilon^{\tau} = \frac{1}{4}\int d^{4}x\sqrt{g}\Big[W^{\alpha\mu,\beta\nu}\mathcal{L_{\epsilon}}\mathcal{\bar{R}}_{\alpha\mu,\beta\nu} \nonumber  \\
  +W^{\alpha\mu,\beta\nu}\mathcal{G}^{\rho}_{\alpha\beta}R_{\rho\mu,\tau\nu}\epsilon^{\tau}- \frac{1}{2}C^{\beta\nu,\alpha}\mathcal{\bar{R}}_{\beta\nu,\alpha\tau}\epsilon^{\tau}
  -\frac{1}{2}W^{\alpha\mu\beta\nu}R_{\rho\mu,\beta\nu}\mathcal{G}^{\rho}_{\alpha\tau}\epsilon^{\tau}\Big]. \label{1.48}
\end{align}
Now we are good prepared to construct gauge and Weyl invariant action for spin 2 case. For that we should calculate gauge variation of Weyl invariant Lagrangian (\ref{1.18}). Using (\ref{1.44}) and
\begin{equation}\label{1.49}
  \bar{\delta}\mathcal{G}_{\tau;\mu\nu}=\nabla_{\mu}\nabla_{\nu}\epsilon_{\tau}-R_{\nu\tau,\mu\rho}\epsilon^{\rho},
\end{equation}
we obtain
\begin{align}
  \bar{\delta}S_{\mathcal{G}}=\int d^{4}x\sqrt{g}W^{\alpha\mu,\beta\nu}\Big(\mathcal{G}^{\tau}_{\alpha\beta}
  (\nabla_{\mu}\nabla_{\nu}\epsilon_{\tau}-R_{\nu\tau,\mu\rho}\epsilon^{\rho})
 -\bar{\delta}h_{\alpha\beta}\mathcal{K}_{\mu\nu}\nonumber\\-h_{\alpha\beta}\mathcal{L_{\epsilon}}K_{\mu\nu}
 -\frac{1}{4}h_{\alpha\beta}W_{\tau\mu,\rho\nu}h^{\tau\rho} \Big), \label{1.50}
\end{align}
where
\begin{equation}
  S_{\mathcal{G}}=\int d^{4}x\sqrt{g}\mathcal{L}^{W\mathcal{G}^{2}}_{-4} .
\end{equation}
Then doing partial integration and using relations (\ref{1.38}) and (\ref{1.48}) we arrive to the following intermediate formula:
\begin{align}
  \bar{\delta}S_{\mathcal{G}}&=\int d^{4}x\sqrt{g}\Big(-\frac{1}{2}B^{\alpha\beta}\mathcal{L_{\epsilon}}h_{\alpha\beta}-\frac{1}{16}  W^{\alpha\mu,\beta\nu}W_{\alpha\mu,\beta\nu}(\mathcal{L_{\epsilon}}h^{\tau}_{\tau}+\frac{1}{2}\bar{\delta}
  [h_{\rho\tau}h^{\rho\tau}])\nonumber\\
  &-\frac{1}{8}\bar{\delta}(W^{\alpha\mu,\beta\nu}h_{\mu\nu}W_{\alpha\rho,\beta\tau}h^{\rho\tau})
  +\frac{1}{4}W^{\alpha\mu,\beta\nu}\mathcal{L_{\epsilon}}\mathcal{W}_{\alpha\mu,\beta\nu}
  +\frac{1}{4}W^{\alpha\mu,\beta\nu}\mathcal{L_{\epsilon}}[W_{\alpha\mu,\beta\tau}h^{\tau}_{\nu}] \Big) .\label{1.51}
\end{align}
To proceed more we should widely use relations:
\begin{eqnarray}
  \bar{\delta}h_{\alpha\beta} &=& \mathcal{L_{\epsilon}}g_{\alpha\beta}, \label{1.52}\\
   \bar{\delta}h^{\alpha\beta} &=& -\mathcal{L_{\epsilon}}g^{\alpha\beta} , \label{1.53}
\end{eqnarray}
and (\ref{1.55})-(\ref{1.57}).
Using these we arrive to the following variation:
\begin{align}
  & \bar{\delta}\Big(S_{\mathcal{G}}+\frac{1}{32}\int d^{4}x\sqrt{g}W^{\alpha\mu,\beta\nu}
   W_{\alpha\mu,\beta\nu}[h_{\rho\tau}h^{\rho\tau}-\frac{1}{4}h^{\rho}_{\rho}h^{\tau}_{\tau}]\Big)\nonumber\\
  & =\int d^{4}x\sqrt{g}\Big\{ -\frac{1}{2}B^{\alpha\beta}\mathcal{L_{\epsilon}}h_{\alpha\beta}-\frac{1}{32}W^{\alpha\mu,\beta\nu}
   W_{\alpha\mu,\beta\nu}\mathcal{L_{\epsilon}}h^{\rho}_{\rho}+\frac{1}{4}W^{\alpha\mu,\beta\nu}
   \mathcal{L_{\epsilon}}\mathcal{W}_{\alpha\mu,\beta\nu}\Big\}. \label{1.58}
\end{align}
Then investigating gauge variation of another Weyl invariant action (\ref{1.21}) we derive:
\begin{align}
  &\frac{1}{4}\bar{\delta}\Big(S_{\mathcal{W}} +\frac{1}{4}\int d^{4}x\sqrt{g}\Big\{\mathcal{W}^{\alpha\mu,\beta\nu}W_{\alpha\mu,\beta\nu}h^{\rho}_{\rho}
  +\frac{1}{8}W^{\alpha\mu,\beta\nu}W_{\alpha\mu,\beta\nu}h^{\rho}_{\rho}h^{\tau}_{\tau}\Big\}\Big) \nonumber\\
  &=\int d^{4}x\sqrt{g}\Big\{\frac{1}{4}\mathcal{W}^{\alpha\mu,\beta\nu}\mathcal{L_{\epsilon}}W_{\alpha\mu,\beta\nu}
  +\frac{1}{16}h^{\rho}_{\rho}W^{\alpha\mu,\beta\nu}\mathcal{L_{\epsilon}}W_{\alpha\mu,\beta\nu}\Big\},\label{1.59}
\end{align}
where
\begin{equation}
  S_{\mathcal{W}}=\int d^{4}x\sqrt{g}\mathcal{W}^{\alpha\mu,\beta\nu}\mathcal{W}_{\alpha\mu,\beta\nu}.
\end{equation}
Summing (\ref{1.58}) and (\ref{1.59}) we obtain
\begin{align}
  & \bar{\delta}\Big(S_{\mathcal{G}}+\frac{1}{4}S_{\mathcal{W}}+ \frac{1}{16}\int d^{4}x\sqrt{g}\Big\{\mathcal{W}^{\alpha\mu,\beta\nu}W_{\alpha\mu,\beta\nu}h^{\rho}_{\rho}+\frac{1}{2}W^{\alpha\mu,\beta\nu}
   W_{\alpha\mu,\beta\nu}h_{\rho\tau}h^{\rho\tau}\Big\}\Big)\nonumber\\
  & =\int d^{4}x\sqrt{g}\Big\{\frac{1}{4}(W^{\alpha\mu,\beta\nu}\mathcal{L_{\epsilon}}\mathcal{W}_{\alpha\mu,\beta\nu}
  +\mathcal{W}^{\alpha\mu,\beta\nu}\mathcal{L_{\epsilon}}W_{\alpha\mu,\beta\nu})\nonumber\\
  &+\frac{1}{16}h^{\rho}_{\rho}W^{\alpha\mu,\beta\nu}\mathcal{L_{\epsilon}}W_{\alpha\mu,\beta\nu}
  -\frac{1}{32}W^{\alpha\mu,\beta\nu}W_{\alpha\mu,\beta\nu}\mathcal{L_{\epsilon}}h^{\rho}_{\rho}
   -\frac{1}{2}B^{\alpha\beta}\mathcal{L_{\epsilon}}h_{\alpha\beta}\Big\}. \label{1.60}
\end{align}
Finally investigating first three terms of r.h.s. of (\ref{1.60}) we see due to (\ref{1.52})-(\ref{1.56}) that
\begin{align}
  &\int d^{4}x\sqrt{g}\Big\{\frac{1}{4}(W^{\alpha\mu,\beta\nu}\mathcal{L_{\epsilon}}\mathcal{W}_{\alpha\mu,\beta\nu}
  +\mathcal{W}^{\alpha\mu,\beta\nu}\mathcal{L_{\epsilon}}W_{\alpha\mu,\beta\nu})
  +\frac{1}{16}h^{\rho}_{\rho}W^{\alpha\mu,\beta\nu}\mathcal{L_{\epsilon}}W_{\alpha\mu,\beta\nu}\Big\}\nonumber \\
  &=\frac{1}{8}\bar{\delta}\int d^{4}x\sqrt{g}\Big\{\mathcal{W}^{\alpha\mu,\beta\nu}W_{\alpha\mu,\beta\nu}h^{\rho}_{\rho}
  +\frac{1}{16}W^{\alpha\mu,\beta\nu}W_{\alpha\mu,\beta\nu}h^{\rho}_{\rho}h^{\tau}_{\tau}\Big\}\nonumber\\
  &+\frac{1}{32}\int d^{4}x\sqrt{g}W^{\alpha\mu,\beta\nu}W_{\alpha\mu,\beta\nu}\mathcal{L_{\epsilon}}h^{\rho}_{\rho}.\label{1.62}
\end{align}
Combining with (\ref{1.60}) we see that nonintegrable terms with $\mathcal{L_{\epsilon}}h^{\rho}_{\rho}$ cancel together and we arrive to the following final formula:
\begin{align}
  & \bar{\delta}\Big(S_{\mathcal{G}}+\frac{1}{4}S_{\mathcal{W}}- \frac{1}{16}\int d^{4}x\sqrt{g}\Big\{\mathcal{W}^{\alpha\mu,\beta\nu}W_{\alpha\mu,\beta\nu}h^{\rho}_{\rho}-\frac{1}{2}W^{\alpha\mu,\beta\nu}
   W_{\alpha\mu,\beta\nu}[h_{\rho\tau}h^{\rho\tau}-\frac{1}{4}h^{\rho}_{\rho}h^{\tau}_{\tau}]\Big\}\Big)\nonumber\\
  & =\int d^{4}x\sqrt{g}\Big\{-\frac{1}{2}B^{\alpha\beta}\mathcal{L_{\epsilon}}h_{\alpha\beta}\Big\}. \label{1.62}
\end{align}
Therefore we prove that in the background with zero Bach tensor
gauge and Weyl invariant action is (\ref{1.64})

\section{Appendix C. Spin Three Details}
\setcounter{equation}{0}
\renewcommand{\theequation}{C.\arabic{equation}}
\quad
Here we present the Weyl transformations of first two  Christoffel symbols and  corresponding traceless parts of them. For first Christoffel symbol transformation looks like:
\begin{align}
  &(\delta-4\sigma) \Gamma^{(1)}_{\gamma;\mu\nu\lambda}= 4 \sigma_{\gamma} h_{\mu\nu\lambda} -2 g_{(\mu\nu}h_{\lambda)\gamma\tau}\sigma^{\tau}, \label{3.55}\\
  &(\delta-6\sigma) \Gamma^{(1)\quad \mu}_{\gamma;\lambda\mu}= 4 \sigma_{\gamma} h^{\mu}_{\mu\lambda} -12 h_{\lambda\gamma\tau}\sigma^{\tau},\label{3.56}\\
  & (\delta-6\sigma) \Gamma^{(1)\quad \gamma}_{\gamma;\nu\lambda}=  -2g_{\nu\lambda} h^{\mu}_{\mu\tau}\sigma^{\tau},\label{3.57}\\
  & (\delta - 4\sigma)\Gamma^{(1);T}_{\gamma;\mu\nu\lambda}=
  4\sigma_{\gamma}h^{T}_{\mu\nu\lambda},\label{3.58}\\
& (\delta - 6\sigma)\Gamma^{(1);T\quad \gamma}_{\gamma;\nu\lambda}=4h^{T}_{\nu\lambda\tau}\sigma^{\tau},\label{3.59}
\end{align}
then introducing notations
\begin{eqnarray}
  t_{\mu\nu\lambda} &=& \sigma^{\tau}\Gamma^{(1)}_{\tau;\mu\nu\lambda}, \label{3.60}\\
  t_{\lambda} &=& t^{\mu}_{\mu\lambda}, \label{3.61}\\
  t^{T}_{\mu\nu\lambda}&=& t_{\mu\nu\lambda}-\frac{1}{6}g_{(\mu\nu}t_{\lambda)}, \label{3.62}\\
  \Gamma^{T}_{\mu\nu\lambda}&=& \Gamma^{(2);T\alpha}_{\alpha\quad\,\,\,\,;\mu\nu\lambda},
\end{eqnarray}
we collect all formulas for second Christoffel symbol:
\begin{eqnarray}
(\delta-4\sigma) \Gamma^{(2)}_{\beta\gamma;\mu\nu\lambda} &=& 3 \sigma_{(\beta}\Gamma^{(1)}_{\gamma);\mu\nu\lambda}\nonumber\\
 &+&2 g_{\beta\gamma}t_{\mu\nu\lambda} - g_{(\mu<\beta}t_{\gamma>\nu\lambda)}
 +2 g_{(\mu\nu}t_{\lambda)\beta\gamma}\nonumber\\
 &+&6g_{(\mu\nu}\nabla_{\lambda)}h_{\beta\gamma\tau}\sigma^{\tau},\label{3.63}\\
  (\delta-2\sigma) \Gamma^{(2)\quad \alpha}_{\beta\gamma;\lambda\alpha} &=& 3\sigma_{(\beta}\Gamma^{(1)\quad\alpha}_{\gamma);\lambda\alpha}+8t_{\gamma\beta\lambda}
  +2g_{\beta\gamma}t_{\lambda}-g_{\lambda(\beta}t_{\gamma)}\nonumber\\
  &+&36\nabla_{\lambda}h_{\beta\gamma\tau}\sigma^{\tau} ,\label{3.64}\\
 (\delta-2\sigma) \Gamma^{(2)\quad \alpha}_{\beta[\gamma;\lambda]\alpha} &=& 3\sigma_{\beta}\Gamma^{(1)\quad\alpha}_{[\gamma;\lambda]\alpha}
 +3\sigma_{[\gamma}\Gamma^{(1)\quad\alpha}_{\beta;\lambda]\alpha}\nonumber\\
 &+&3g_{\beta[\gamma}t_{\lambda]}-18\Gamma^{(1)}_{[\gamma
 ;\lambda]\beta\tau}\sigma^{\tau},\label{3.65}\\
(\delta-2\sigma) \Gamma^{(2)\quad \alpha}_{\beta\alpha;\nu\lambda} &=& 3\sigma_{\beta}\Gamma^{(1)\quad\alpha}_{\alpha;\nu\lambda}-4t_{\beta\nu\lambda}\nonumber\\
  &-&\frac{3}{2}g_{\nu\lambda}\Gamma^{(1)\quad\alpha}_{\beta;\alpha\tau}\sigma^{\tau}
  +\frac{1}{2}g_{\nu\lambda}t_{\beta}-g_{\beta(\nu}t_{\lambda)},\label{3.66}\\
(\delta-4\sigma) \Gamma^{(2);T}_{\beta\gamma;\mu\nu\lambda} &=& 3 \sigma_{(\beta}\Gamma^{(1);T}_{\gamma);\mu\nu\lambda}
+2g_{\beta\gamma}t^{T}_{\mu\nu\lambda}\nonumber\\
&-& g_{(\mu<\beta}t_{\gamma>\nu\lambda)}
 +\frac{2}{3}g_{(\mu\nu}t_{\lambda)\beta\gamma}
 +\frac{1}{6}g_{(\mu\nu}g_{\lambda)(\beta}t_{\gamma)},\label{3.67}\\
 (\delta-2\sigma) \Gamma^{T}_{\mu\nu\lambda}  &=& 8t^{T}_{\mu\nu\lambda},\label{3.68}\\
 (\delta-2\sigma) \Gamma^{(2);T\quad \alpha}_{\beta\alpha;\nu\lambda} &=& 3\sigma_{\beta}\Gamma^{(1);T\,\,\alpha}_{\alpha;\nu\lambda}+3\Gamma^{(1);T}_{\beta;\nu\lambda\tau}\sigma^{\tau}
 -\frac{11}{3}t^{T}_{\beta\nu\lambda}\nonumber\\
  &+&\frac{8}{9}\big(g_{\nu\lambda}t_{\beta}
  -2g_{\beta(\nu}t_{\lambda)}\big)=(\delta-6\sigma)\gamma^{(2);T}_{\beta;\nu\lambda},\label{3.69}\\
 (\delta-4\sigma) \Gamma^{(2)T;T}_{\beta[\gamma;\lambda]\mu\nu} &=& (\delta-4\sigma)\Big[ \Gamma^{(2);T}_{\beta[\gamma;\lambda]\mu\nu}
 -\frac{1}{4}g_{[\gamma(\mu}\gamma^{(2);T}_{\beta;\nu)\lambda]}\Big]\nonumber\\
 &=&3\sigma_{\beta}\Big(\Gamma^{(1);T}_{[\gamma;\lambda]\mu\nu}
 -\frac{1}{4}g_{[\gamma(\mu}\Gamma^{(1);T\quad\alpha}_{|\alpha|;\nu)\lambda]}\Big)\nonumber\\
 &+&3\Big(\sigma_{[\gamma}\Gamma^{(1);T}_{|\beta|;\lambda]\mu\nu}
 -\frac{1}{4}g_{[\gamma(\mu}\Gamma^{(1);T}_{|\beta|;\nu)\lambda]\tau}\sigma^{\tau}\Big)\nonumber\\
&+&3\Big(g_{\beta[\gamma}t^{T}_{\lambda]\mu\nu}
 -\frac{1}{4}g_{[\gamma(\mu}t^{T}_{\nu)\lambda]\beta}\Big).\label{3.70}
\end{eqnarray}

We see that r.h.s of latter consists of three brackets and each of them is just traceless part of first term. To see cancelation of odd terms in this variation we present for completeness two more useful variations:
\begin{align}
 (\delta-4\sigma) \Gamma^{(2);T}_{\beta[\gamma;\lambda]\mu\nu} &= 3\sigma_{\beta}\Gamma^{(1);T}_{[\gamma;\lambda]\mu\nu}
 +3\sigma_{[\gamma}\Gamma^{(1);T}_{\beta;\lambda]\mu\nu}
 +3g_{\beta[\gamma}t^{T}_{\lambda]\mu\nu}
 -\frac{5}{3}g_{[\gamma(\mu}t^{T}_{\nu)\lambda]\beta}\nonumber\\
   &+\frac{4}{9}\big( g_{\beta[\gamma}g_{\lambda](\mu}t_{\nu)}-  g_{\beta(\mu}g_{\nu)[\gamma}t_{\lambda]}\big) , \nonumber\\
   (\delta-4\sigma)\big[-\frac{1}{4}
   g_{[\gamma(\mu}\gamma^{(2);T}_{\beta;\nu)\lambda]}\big]&=
   -\frac{3}{4}\sigma_{\beta}g_{[\gamma(\mu}\Gamma^{(1);T\quad\alpha}_{|\alpha|;\nu)\lambda]}
   -\frac{3}{4}g_{[\gamma(\mu}
   \Gamma^{(1);T}_{|\beta|;\nu)\lambda]\tau}\sigma^{\tau}
   +\frac{11}{12}g_{[\gamma(\mu}t^{T}_{\nu)\lambda]\beta}\nonumber\\
   &-\frac{4}{9}\big( g_{\beta[\gamma}g_{\lambda](\mu}t_{\nu)}-  g_{\beta(\mu}g_{\nu)[\gamma}t_{\lambda]}\big).\nonumber
\end{align}
Summing last two variations we restore (\ref{3.70}).

The next task is to construct analog of (\ref{1.18}) in spin three case.
For this we need to reconsider second Weyl invariant for spin three with Weyl tensor, one derivative and linearized spin 3 field.
 In this case we start from the same formula (\ref{4.1}) and replace spin three field with corresponding traceless part:
\begin{eqnarray}
  &&(\delta-4\sigma) [\nabla_{(\alpha}W^{\quad \rho}_{\beta\quad , \gamma)(\mu}h^{T}_{\nu\lambda)\rho}-C_{<\mu(\alpha,\beta}h^{T}_{\gamma)\nu\lambda>}+C^{\quad \rho}_{(\alpha\quad ,\beta}g_{\gamma)(\mu}h^{T}_{\nu\lambda)\rho} \nonumber\\
  &&-2g_{(\alpha\beta}C^{\qquad \rho}_{\gamma)(\mu ,}h^{T}_{\nu\lambda)\rho}-2g_{(\alpha\beta}C^{\quad \rho}_{\gamma)\quad,(\mu }h^{T}_{\nu\lambda)\rho}] = -4\sigma_{(\alpha}W^{\quad \rho}_{\beta\quad,\gamma)(\mu}h^{T}_{\nu\lambda)\rho}.\quad\quad\label{3.74}
\end{eqnarray}
Then using again (\ref{3.58}) we obtain simply:
\begin{equation}\label{3.75} (\delta-4\sigma)[W^{\quad\rho}_{(<\alpha\quad,\beta(\mu}\Gamma^{(1);T}_{\gamma>;\nu\lambda)\rho}]
=4\sigma_{(\alpha}W^{\quad \rho}_{\beta\quad,\gamma)(\mu}h^{T}_{\nu\lambda)\rho}.
\end{equation}
So we see that summing (\ref{3.74}) and (\ref{3.75}) we obtain exact Weyl invariant tensor with two set of symmetrized indices $(\alpha\beta\gamma)$ and $(\mu\nu\lambda)$
\begin{eqnarray}
  T_{\alpha\beta\gamma,\mu\nu\lambda} &=& \nabla_{(\alpha}W^{\quad \rho}_{\beta\quad , \gamma)(\mu}h^{T}_{\nu\lambda)\rho}
  -C_{<\mu(\alpha,\beta}h^{T}_{\gamma)\nu\lambda>}+C^{\quad \rho}_{(\alpha\quad ,\beta}g_{\gamma)(\mu}h^{T}_{\nu\lambda)\rho}\nonumber\\
  &&-2g_{(\alpha\beta}C^{\qquad \rho}_{\gamma)(\mu ,}h^{T}_{\nu\lambda)\rho}-2g_{(\alpha\beta}C^{\quad \rho}_{\gamma)\quad,(\mu }h^{T}_{\nu\lambda)\rho}+W^{\quad \rho}_{(<\alpha\quad , \beta(\mu}\Gamma^{(1);T}_{\gamma>;\nu\lambda)\rho}.\qquad\quad\label{3.76}
\end{eqnarray}
Then after antisymmetrization of pairs $[\alpha \mu], [\beta \nu], [\gamma \lambda]$ we have
\begin{align}
 T_{\alpha\mu,\beta\nu,\gamma\rho}&= 4\nabla_{<\alpha}W^{\quad\quad \rho}_{\beta\nu, [\gamma}h^{T}_{\lambda]\mu>\rho}+4\nabla_{<\alpha}W^{\quad\quad \rho}_{\gamma\lambda, [\beta}h^{T}_{\nu]\mu>\rho}\nonumber\\
&+4\nabla_{<\beta}W^{\quad\quad \rho}_{\alpha\mu, [\gamma}h^{T}_{\lambda]\nu>\rho}+4\nabla_{<\beta}W^{\quad\quad \rho}_{\gamma\lambda, [\alpha}h^{T}_{\mu]\nu>\rho}\nonumber\\
  &+4\nabla_{<\gamma}W^{\quad\quad \rho}_{\beta\nu, [\alpha}h^{T}_{\mu]\lambda>\rho}+4\nabla_{<\gamma}W^{\quad\quad \rho}_{\alpha\mu, [\beta}h^{T}_{\nu]\mu>\rho}\nonumber\\
  &-8C^{\quad\rho}_{\gamma\lambda,}g_{<\alpha[\beta}h^{T}_{\mu>\nu]\rho}
  -4g_{<\alpha\{\beta}C^{\quad\rho}_{[\gamma\quad,(\mu>}h^{T}_{\nu\})\lambda]\rho}\nonumber\\
  &-8C^{\quad\rho}_{\alpha\mu,}g_{<\beta[\gamma}h^{T}_{\nu>\lambda]\rho}
  -4g_{<\beta\{\gamma}C^{\quad\rho}_{[\alpha\quad,(\nu>}h^{T}_{\lambda\})\mu]\rho}\nonumber\\
  &-8C^{\quad\rho}_{\beta\nu,}g_{<\gamma[\alpha}h^{T}_{\lambda>\mu]\rho}
  -4g_{<\gamma\{\alpha}C^{\quad\rho}_{[\beta\quad,(\mu>}h^{T}_{\lambda\})\nu]\rho}\nonumber\\
 & -W^{\quad\rho}_{(<\alpha\quad,\{\beta)[\gamma} \Gamma^{(1)T}_{\lambda];\mu>\nu\}\rho} + 3 W^{\qquad\rho}_{\gamma\lambda,(<\beta}\Gamma^{(1)T}_{[\alpha);\mu]\nu>\rho}\nonumber\\
 & -W^{\quad\rho}_{(<\beta\quad,\{\gamma)[\alpha} \Gamma^{(1)T}_{\mu];\nu>\lambda\}\rho} + 3 W^{\qquad\rho}_{\alpha\mu,(<\beta}\Gamma^{(1)T}_{[\gamma);\nu>\lambda]\rho}\nonumber\\
 & -W^{\quad\rho}_{(<\gamma\quad,\{\alpha)[\beta} \Gamma^{(1)T}_{\nu];\lambda>\mu\}\rho} + 3 W^{\qquad\rho}_{\beta\nu,(<\alpha}\Gamma^{(1)T}_{[\gamma);\lambda]\mu>\rho}.\label{3.77}
\end{align}
This is Weyl invariant combination of Weyl tensor, spin 3 field and one covariant derivative with indices organized in a way of Riemann tensor, therefore it is spin 3 version of
\begin{equation}\label{3.78}
  \frac{1}{2}[W_{\alpha\mu,[\beta}^{\qquad\tau} h_{\nu]\tau} + W_{\beta\nu,[\alpha}^{\qquad\tau}h_{\mu]\tau}],
\end{equation}
coming from linearization (\ref{1.46}) of gravitational Riemann tensor. So we can couple (\ref{3.77}) with spin three Weyl tensor and obtain primary:
\begin{eqnarray}
  L_{\mathcal{W}T} &=& \mathcal{W}^{\alpha\mu,\beta\nu,\gamma\lambda}T_{\alpha\mu,\beta\nu,\gamma\lambda}.
\end{eqnarray}
Trace of (\ref{3.78})  is traceless symmetric tensor $W^{\alpha\mu,\beta\nu}h_{\mu\nu}$
contracted with Schouten tensor $K_{\alpha\beta}$ in the last term of Weyl invariant (\ref{1.18}). Therefore we can try to construct something similar in spin three case.
For this purpose we can calculate trace of invariant (\ref{3.77})
\begin{align}
  T_{\mu\nu;\gamma\lambda} & = g^{\alpha\beta}T_{\alpha\mu,\beta\nu,\gamma\rho}=4 \nabla_{[\gamma}W^{\quad\tau\quad\rho}_{(\mu\quad,\nu)}h^{T}_{\lambda]\tau\rho}
  + 4 \nabla_{(\mu}W^{\quad\tau\quad\rho}_{\nu)\quad,[\gamma}h^{T}_{\lambda]\tau\rho}
  +4\nabla_{[\gamma}W^{\quad\tau\quad\rho}_{\lambda]\quad,(\mu}h^{T}_{\nu)\tau\rho}\nonumber\\
&+ W^{\quad\tau\quad\rho}_{[\gamma\quad,(\mu}\Gamma^{(1)T}_{[\nu);\lambda]]\tau\rho}
+2W^{\quad(\tau\quad\rho)}_{[\gamma\quad,(\mu}\Gamma^{(1)T}_{[\nu);\tau]\lambda]\rho}
+3W^{\quad\tau\quad\rho}_{(\mu\quad,\nu)}\Gamma^{(1)T}_{[\gamma;\lambda]\tau\rho}\nonumber\\
& -8C^{\quad\rho}_{[\gamma\quad,(\mu}h^{T}_{\nu)\lambda]\rho} - 8C^{\quad\rho}_{(\mu\quad,[\gamma}h^{T}_{\nu)\lambda]\rho}
-W^{\qquad\rho}_{[\gamma(\mu,\nu)}\Gamma^{(1)T\,\,\,\,\tau}_{\tau;\lambda]\rho}
-3W^{\qquad\rho}_{\gamma\lambda,(\mu}\Gamma^{(1)T\,\,\,\,\tau}_{\tau;\nu)\rho}\nonumber\\
&+6W^{\quad\tau\rho}_{\gamma\lambda,}\Gamma^{(1)T}_{\tau;\rho\mu\nu}
-24C^{\quad\rho}_{\gamma\lambda,}h^{T}_{\rho\mu\nu} +3W^{\quad\,\,\,\,\tau\rho}_{[\gamma(\mu,}\Gamma^{(1)T}_{\tau;\rho\nu)\lambda]}
-12C^{\quad\,\,\,\,\rho}_{[\gamma(\mu,}h^{T}_{\nu)\lambda]\rho}\nonumber\\
&-8g_{\mu\nu}C^{\quad\tau\,\,\,\,\rho}_{[\gamma\quad,}h^{T}_{\lambda]\tau\rho}
-4g_{[\gamma(\mu}C^{\quad\tau\,\,\,\,\rho}_{\lambda]\quad,}h^{T}_{\nu)\tau\rho}
-12g_{[\gamma(\mu}C^{\quad\tau\,\,\,\,\rho}_{\nu)\quad,}h^{T}_{\lambda]\tau\rho}.\label{3.83}
\end{align}
Now we see that the last expression can be split up to the six primary fields:
\begin{align}
I^{(1)}_{\mu\nu;\gamma\lambda}&=4\nabla_{[\gamma}W_{(\mu \ , \nu)}^{\ \ \tau \ \ \rho}h_{\lambda]}^{T\tau \rho}+
2W_{(\mu \ , \nu)}^{\ \ \tau \ \ \rho}\Gamma^{(1)T}_{[\gamma,\lambda]\tau\rho}+
2W_{[\gamma \ , (\mu}^{\ \ \tau \ \ \rho}\Gamma^{(1)T}_{\nu),\lambda]\tau \rho}-
2W_{[\gamma(\mu,\nu)}^{\qquad  \rho}\Gamma_{T \ \ ;\tau\lambda]\rho}^{(1) \tau}\nonumber\\&-
8C_{(\mu \, ;}^{\ \, \rho \ \tau}g_{\nu)[\gamma}h^{T}_{\lambda] \tau \rho},\\
I^{(2)}_{\mu\nu;\gamma\lambda}&=4\nabla_{(\mu}W_{\nu) \ , [\gamma}^{\ \ \tau \ \ \rho}h_{\lambda]\tau \rho}+
W_{(\mu \ , \nu)}^{\ \ \tau \ \ \rho}\Gamma^{(1)T}_{[\gamma,\lambda]\tau\rho}+
3W_{[\gamma \ , (\mu}^{\ \ \tau \ \ \rho}\Gamma^{(1)T}_{\nu),\lambda]\tau \rho}+
W_{[\gamma(\mu,\nu)}^{\qquad  \rho}\Gamma_{T \ \ ;\tau\lambda]\rho}^{(1) \tau}\nonumber \\&+4(C_{[\gamma \ \, ;(\mu}^{\ \ \rho}+C_{(\mu \ \, ;[\gamma}^{\ \ \rho})h^{T}_{\lambda] \nu) \rho}-
4C_{(\mu \ ;}^{\ \,\, \rho \ \tau}g_{\nu)[\gamma}h^{T}_{\lambda] \tau \rho}-
8g_{\mu \nu}C_{[\gamma \, ;}^{\ \, \rho \ \tau}h^{T}_{\lambda]\rho \tau},\\
I^{(3)}_{\mu\nu;\gamma\lambda}&=4\nabla_{[\gamma}W_{\lambda] \ , (\mu}^{\ \ \tau \ \ \rho}h_{\nu)\tau \rho}-
W_{[\gamma \ , (\mu}^{\ \ \tau \ \ \rho}\Gamma^{(1)T}_{\lambda],\nu)\tau \rho}-
3W_{\gamma\lambda,(\mu}^{\qquad  \rho}\Gamma_{T \ \ ;\tau\nu)\rho}^{(1) \tau}\nonumber\\&-
4(C_{[\gamma \ \, ;(\mu}^{\ \ \rho}+C_{(\mu \ \, ;[\gamma}^{\ \ \rho})h^{T}_{\lambda] \nu) \rho}+
4C_{[\gamma \, ;}^{\ \, \tau \ \rho}g_{\lambda](\mu}h^{T}_{\nu) \tau \rho},\\
 I^{(4)}_{\mu\nu;\gamma\lambda}&= 6W^{\quad\tau\rho}_{\gamma\lambda,}\Gamma^{(1)T}_{\tau;\rho\mu\nu}
-24C^{\quad\rho}_{\gamma\lambda,}h^{T}_{\rho\mu\nu},\label{3.87}\\
 I^{(5)}_{\mu\nu;\gamma\lambda}&= 3W^{\quad\tau\rho}_{[\gamma(\mu,}\Gamma^{(1)T}_{\tau;\rho\nu)\lambda]}
-12C^{\quad\rho}_{[\gamma,(\mu}h^{T}_{\nu)\lambda]\rho},\\
 I^{(6)}_{\mu\nu;\gamma\lambda}&= -2W^{\quad(\tau\quad\rho)}_{[\gamma\quad,(\mu}\Gamma^{(1)T}_{\tau;\rho\nu)\lambda]}
  -8(C_{[\gamma \ \, ;(\mu}^{\ \ \rho}+C_{(\mu \ \, ;[\gamma}^{\ \ \rho})h^{T}_{\lambda] \nu) \rho}.
\end{align}
Then first of all we see that
\begin{equation}
  \sum^{6}_{ i=1}I^{(i)}_{\mu\nu;\gamma\lambda}=T_{\mu\nu;\gamma\lambda}.
\end{equation}
Moreover contracting with spin three Schouten tensor $K_{\mu\nu;\gamma\lambda}$ and
using Bianchi identity we arrive to the idea that under Bianchi projection the sum of second and third invariant is equal to first:
\begin{equation}
  I^{(1)}_{\mu\nu;\gamma\lambda}K^{\mu\nu;\gamma\lambda}=
  (I^{(2)}_{\mu\nu;\gamma\lambda}+I^{(3)}_{\mu\nu;\gamma\lambda})K^{\mu\nu;\gamma\lambda}.
 \end{equation}
 Then fifth invariant is also not independent and connected with fourtn.
 \begin{equation}
   I^{(5)}_{\mu\nu;\gamma\lambda}=-\frac{1}{2}I^{(4)}_{(\mu[\gamma;\lambda]\nu)}.
 \end{equation}
Therefore we have only 4 independent invariants $I^{(1)}_{\mu\nu;\gamma\lambda}, I^{(2)}_{\mu\nu;\gamma\lambda}, I^{(4)}_{\mu\nu;\gamma\lambda}$ and $I^{(6)}_{\mu\nu;\gamma\lambda}$:
The last one can be combined with first two to obtain invariant expression with Christoffel symbols with one symmetric and one antisymmetric pair of indices. This is necessary for integration to the square of second Christoffel symbols with the similar organization of indices in the future construction.
\begin{align}
  J^{(1)}_{\mu\nu;\gamma\lambda} &= I^{(1)}_{\mu\nu;\gamma\lambda}+\frac{1}{2}I^{(6)}_{\mu\nu;\gamma\lambda}\nonumber\\
  &=4\nabla_{[\gamma}W_{(\mu \ , \nu)}^{\ \ \tau \ \ \rho}h_{\lambda]}^{\tau \rho}+
2W_{(\mu \ , \nu)}^{\ \ \tau \ \ \rho}\Gamma^{(1)T}_{[\gamma,\lambda]\tau\rho}+
W_{[\gamma \ , (\mu}^{\ \ (\tau \ \ \rho)}\big[\Gamma^{(1)T}_{\nu),\tau\lambda] \rho}-\Gamma^{(1)T}_{\tau,\nu)\lambda] \rho}\big]
\nonumber\\&-2W_{[\gamma(\mu,\nu)}^{\qquad  \rho}\Gamma_{T \ \ ;\tau\lambda]\rho}^{(1) \tau}-4(C_{[\gamma \ \, ;(\mu}^{\ \ \rho}+C_{(\mu \ \, ;[\gamma}^{\ \ \rho})h^{T}_{\lambda] \nu) \rho}-
8C_{(\mu \, ;}^{\ \, \rho \ \tau}g_{\nu)[\gamma}h^{T}_{\lambda] \tau \rho},
\end{align}
\begin{align}
  J^{(2)}_{\mu\nu;\gamma\lambda} &= I^{(2)}_{\mu\nu;\gamma\lambda}+\frac{3}{4}I^{(6)}_{\mu\nu;\gamma\lambda}\nonumber\\
  &=4\nabla_{(\mu}W_{\nu) \ , [\gamma}^{\ \ \tau \ \ \rho}h_{\lambda]\tau \rho}+
W_{(\mu \ , \nu)}^{\ \ \tau \ \ \rho}\Gamma^{(1)T}_{[\gamma,\lambda]\tau\rho}+
\frac{3}{2}W_{[\gamma \ , (\mu}^{\ \ (\tau \ \ \rho)}\big[\Gamma^{(1)T}_{\nu),\lambda]\tau \rho}-\Gamma^{(1)T}_{\tau,\nu)\lambda] \rho}\big]\nonumber \\&+
W_{[\gamma(\mu,\nu)}^{\qquad  \rho}\Gamma_{T \ \ ;\tau\lambda]\rho}^{(1) \tau}-2(C_{[\gamma \ \, ;(\mu}^{\ \ \rho}+C_{(\mu \ \, ;[\gamma}^{\ \ \rho})h^{T}_{\lambda] \nu) \rho}-
4C_{(\mu \ ;}^{\ \,\, \rho \ \tau}g_{\nu)[\gamma}h^{T}_{\lambda] \tau \rho}-
8g_{\mu \nu}C_{[\gamma \, ;}^{\ \, \rho \ \tau}h^{T}_{\lambda]\rho \tau}.\label{3.94}
\end{align}
This two primaries we can contract with traceless part of Schouten tensor $K^{(T)}_{\mu\nu;\gamma\lambda}$ with the Weyl transformation (\ref{3.95}). For first invariant tensor $J^{(1)}$ we have:
\begin{align}
  &J^{(1)}_{\mu\nu;\gamma\lambda}K_{(T)}^{\mu\nu;\gamma\lambda}
   = \Big[ 16\nabla_{\gamma}W^{\quad\tau\quad\rho}_{\mu\quad,\nu}
   h^{T}_{\lambda\tau\rho}
   -16C^{\quad\rho}_{(\gamma\quad,\mu)}h^{T}_{\nu\lambda\rho} -4W^{\qquad\rho}_{\gamma\lambda,\mu}\Gamma^{(1)T\,\,\,\,\tau}_{\tau;\nu\rho}\Big]
   K_{(T)}^{\mu\nu;\gamma\lambda}\nonumber\\ & +\Big[4W^{\quad(\tau\quad\rho)}_{\gamma\quad,\mu}
  \Gamma^{(1)T}_{[\nu;\tau]\lambda\rho}
+4W^{\quad\tau\quad\rho}_{\mu\quad,\nu}\Gamma^{(1)T}_{[\gamma;\lambda]\tau\rho}
\Big]K_{(T)}^{\mu\nu;\gamma\lambda}.\label{3.96}
\end{align}
Now we start to analyze last expression.
Considering variation of the proposed part of invariant
\begin{equation}\label{3.97}
 \delta [J^{(1)}_{\mu\nu;\gamma\lambda}K_{(T)}^{\mu\nu;\gamma\lambda}]=J^{(1)}_{\mu\nu;\gamma\lambda}\delta K_{(T)}^{\mu\nu;\gamma\lambda},
\end{equation}
and using (\ref{3.95}) we should try to integrate variation (\ref{3.97}) to terms  second order on $ \Gamma^{(2)T;T}$ and linear in gravitational Weyl tensor. To integrate we should use variation (\ref{3.70}) where unpleasant trace terms appear again. Doing that with help of the following relation
\begin{align}
  (\delta-2\sigma) \nabla_{\gamma}W_{\mu\tau,\nu\rho}& = -2\sigma_{\gamma}W_{\mu\tau,\nu\rho}- \sigma_{[\mu}W_{\gamma\tau],\nu\rho}- \sigma_{[\nu}W_{\mu\tau,\gamma\rho]}\nonumber\\
  &+g_{\gamma[\mu}\delta C_{\nu\rho,\tau]}+g_{\gamma[\nu}\delta C_{\mu\tau,\rho]},\nonumber
\end{align}
and after integration of some second order on Christoffel symbols terms we arrive to the reminder:
\begin{gather}
-\frac{7}{3}W_{\gamma\lambda,\mu}^{\quad \ \rho}\tilde{\gamma}_{\beta,\nu\rho}\Gamma^{\beta[\gamma,\lambda]\mu\nu}_{(2)T,T}+
 4W_{\mu \ \ ,\nu}^{\ \ \tau \ \rho}t^{T}_{\tau \rho \lambda}\Gamma^{(2)\  \ \lambda,\mu\nu\beta}_{T,T\beta}+
\frac{8}{3}W_{\gamma\lambda,\mu}^{\ \ \ \ \rho}t^{T}_{\rho \nu \beta}\Gamma^{\beta[\gamma,\lambda]\mu\nu}_{(2)T,T}-
\frac{2}{3}W_{\gamma\lambda,\beta}^{\ \ \ \ \rho}t^{T}_{\rho \mu\nu }\Gamma^{\beta[\gamma,\lambda]\mu\nu}_{(2)T,T}\nonumber\\
+\frac{4}{3}C_{(\gamma \ ,\mu)}^{\ \ \rho}\Gamma^{(1)T}_{\beta,\nu\lambda\rho}\delta\Gamma^{\beta[\gamma,\lambda]\mu\nu}-
\frac{4}{3}\nabla_{\gamma}W_{\mu \ ,\nu}^{\ \tau \ \rho}\Gamma^{(1)T}_{\beta,\lambda\tau\rho}\delta\Gamma^{\beta[\gamma,\lambda]\mu\nu}_{(2)T,T},\label{3.98}
\end{gather}
where
\begin{gather}
  \tilde{\gamma}_{\beta,\nu\lambda}=\sigma_\beta\Gamma^{(1)T\,\,\, \alpha}_{\alpha;\nu\lambda}+
 \Gamma^{(1)T}_{\beta,\nu\lambda\tau}\sigma^\tau+
 t_{\nu\lambda\beta}^{T}.
\end{gather}
To cancel first term of (\ref{3.98}) we should use the following general relation for remaining invariant (\ref{3.87})
\begin{gather}
\frac{A}{3}\delta [I^{(4)}_{\mu\nu;\gamma\lambda}K_{(T)}^{\mu\nu,\gamma\lambda}]=A [W^{\ \ \ \tau \rho}_{\gamma \lambda,}\Gamma^{(1)T}_{[\tau,\rho]\mu\nu}-8C_{\gamma \lambda,}^{\rho}h_{\rho\mu\nu}]\delta K_{(T)}^{\mu\nu,\gamma\lambda} \nonumber\\
=\delta(\frac{A}{18}W_{\gamma \lambda}^{ \ \ \ \tau \rho} \Gamma^{(2)T,T}_{\beta[\tau,\rho]\mu\nu}\Gamma^{\beta [\gamma,\lambda]\mu\nu}_{(2)T,T})-
\delta(\frac{2A}{3}C_{\gamma \lambda,}^{\ \ \ \rho}\Gamma^{(1)T}_{\beta,\rho\mu\nu}\Gamma^{\beta[\gamma,\lambda]\mu\nu}_{(2)T,T})+
\frac{2A}{3}C_{\gamma\lambda,}^{\ \ \rho}\Gamma^{(1)T}_{\beta,\rho\mu\nu}\delta\Gamma^{\beta[\gamma,\lambda]\mu\nu}_{(2)T,T}\nonumber \\-
\frac{2A}{3}W_{\gamma\lambda,\beta}^{\ \ \ \rho}t^{T}_{\rho \ \mu \nu }\Gamma^{\beta[\gamma,\lambda]\mu\nu}_{(2)T,T}+
\frac{A}{3}W_{\gamma\lambda,\mu}^{\ \ \ \rho}\tilde{\gamma}_{\beta,\nu\rho}\Gamma^{\beta[\gamma,\lambda]\mu\nu}_{(2)T,T}.\label{3.100}
\end{gather}
So we see that taking $A=7$ and adding (\ref{3.100}) to the (\ref{3.98}) we arrive to the some integrated terms (will collect later in general formula for integrated terms) and the following reminder:
\begin{gather}
\frac{8}{3}W_{\gamma\lambda,\mu}^{\ \ \ \ \rho}t^{T}_{\rho \nu \beta}\Gamma^{\beta[\gamma,\lambda]\mu\nu}_{(2)T,T}-
\frac{16}{3}W_{\gamma\lambda,\beta}^{\ \ \ \ \rho}t^{T}_{\rho \mu\nu }\Gamma^{\beta[\gamma,\lambda]\mu\nu}_{(2)T,T}+
 4W_{\mu \ \ ,\nu}^{\ \ \tau \ \rho}t^{T}_{\tau \rho \beta}\Gamma^{(2)\  \ \beta,\mu\nu\alpha}_{T,T\alpha}\nonumber\\
+\frac{4}{3}C_{(\gamma \ ,\mu)}^{\ \ \rho}\Gamma^{(1)T}_{\beta,\nu\lambda\rho}\delta\Gamma_{(2)T;T}^{\beta[\gamma,\lambda]\mu\nu}-
\frac{4}{3}\nabla_{\gamma}W_{\mu \ ,\nu}^{\ \tau \ \rho}\Gamma^{(1)T}_{\beta,\lambda\tau\rho}\delta\Gamma^{\beta[\gamma,\lambda]\mu\nu}_{(2)T,T}+
\frac{14}{3}C_{\gamma\lambda,}^{\ \ \ \rho}\Gamma^{(1)T}_{\beta,\rho\mu\nu}\delta\Gamma^{\beta[\gamma,\lambda]\mu\nu}_{(2)T,T}.\label{3.101}
\end{gather}
To continue we should introduce some definitions
\begin{gather}
 T^{\mu \nu \lambda}_{\alpha \beta \gamma}= \Gamma^{\tau,\mu \nu \lambda}_{(1)T}\Gamma^{(1)T}_{\tau,\alpha\beta\gamma}-\frac{1}{2}(h_{T}^{\mu \nu \lambda}\Gamma^{T}_{\alpha \beta \gamma}+h^{T}_{\alpha \beta \gamma}\Gamma_{T}^{\mu \nu \lambda}),\label{3.106}\\
 T^{\mu \nu }_{\alpha \beta }= T^{\mu \nu \lambda}_{\alpha \beta \lambda}, \quad\quad T^{\mu }_{\alpha}= T^{\mu \nu }_{\alpha \nu }.\label{3.108}
\end{gather}
Note that (\ref{3.106})-(\ref{3.108}) are Weyl invariant tensors.
Then we see that after transformation and integration of some terms in first line of (\ref{3.101}) using (\ref{3.102}) and after applying the following important formula
\begin{gather}
\Gamma_{\beta, a b c}^{(1)T}\delta \Gamma^{\beta[\gamma,\lambda]\mu\nu}_{(2)T,T}=3\sigma^{[\gamma}T^{\lambda]\mu\nu}_{a b c}-
\frac{3}{4}g^{[\gamma(\mu}T^{\nu)\lambda]\tau}_{abc}\sigma_\tau+
\frac{3}{8}\delta[\Gamma^{T}_{abc}\Gamma^{[\gamma,\lambda]\mu\nu}_{(1)T}+\Gamma^{(1)[\gamma}_{T \quad ,abc}\Gamma_{T}^{\lambda]\mu \nu}]\nonumber \\-
\frac{3}{32}\delta[{\Gamma^{T}_{abc}g^{[\gamma(\mu}}\Gamma^{(1)\ \alpha , \nu)\lambda]}_{T\alpha}+\Gamma^{\beta,}_{(1)T\ abc}g^{[\gamma(\mu}\Gamma^{\nu)\lambda]}_{T\quad \ \beta}] \label{1.109}
\end{gather}
to the second line of (\ref{3.101}) we obtain miraculous cancelation of the all overall terms and we arrive to the following reminder expressed through the invariant tensors (\ref{3.106})-(\ref{3.108}):
\begin{gather}
4\sigma^\lambda \nabla_\mu W_{\nu \ ,\gamma}^{\ \tau \ \, \rho}T^{\gamma \mu \nu}_{\tau \rho \lambda}-
4\sigma^\gamma\nabla_{\gamma}W_{\mu \ ,\nu}^{\ \rho \ \, \tau}T^{\mu \nu}_{\rho \tau}+
24\sigma_{\mu}C^{\ \tau,\rho}_{\nu}T^{\mu \nu}_{\tau \rho}+
32\sigma^{\gamma}C_{\gamma \ ,\rho}^{\ \mu}T^{\rho}_{\mu}.\label{3.110}
\end{gather}
Before continue with this reminder we present the integrated terms during the whole procedure presented from formula (\ref{3.97})
\begin{gather}
L^{(1)}=-\frac{2}{9}W_{\mu \ ,\nu}^{\ \tau \ \, \rho}\Gamma^{(2)T,T}_{\beta[\gamma,\lambda]\tau \rho}\Gamma^{\beta[\gamma,\lambda]\mu\nu}_{(2)T,T}
-\frac{2}{9}W_{\gamma \ ,\mu}^{\ (\tau \ \, \rho)}\Gamma^{(2)T,T}_{\beta[\nu,\tau]\lambda \rho}\Gamma^{\beta[\gamma,\lambda]\mu\nu}_{(2)T,T}
-\frac{7}{18}W_{\gamma\lambda,}^{\ \ \tau \rho}\Gamma^{(2)T,T}_{\beta[\tau,\rho]\mu\nu}\Gamma^{\beta[\gamma,\lambda]\mu\nu}_{(2)T,T}\nonumber\\
-\frac{1}{24}W_{\mu \ ,\nu}^{\ \tau \ \, \rho}\Gamma^{T}_{\tau \rho \lambda}\Gamma_{T}^{\mu \nu \lambda}\nonumber\\
-\frac{1}{3}W_{\gamma\lambda,\mu}^{\ \ \ \rho}\Gamma^{T}_{\rho\nu\beta}\Gamma^{\beta[\gamma,\lambda]\mu\nu}_{(2)T,T}+\frac{2}{3}W_{\gamma\lambda,\beta}^{\ \ \ \ \rho}\Gamma^{T}_{\rho\mu\nu}\Gamma^{\beta[\gamma,\lambda]\mu\nu}_{(2)T,T}
-\frac{1}{2}W_{\mu \ \ ,\nu}^{\ \tau \ \ \rho}\Gamma^{T}_{\tau\rho\beta}\Gamma^{(2)\ \beta,\mu\nu\alpha}_{T\alpha}\nonumber \\
-\frac{4}{3}\nabla_{\gamma}W_{\mu \ ,\nu}^{\ \tau \ \, \rho}\Gamma^{(1)T}_{\beta,  \lambda \tau \rho}\Gamma^{\beta[\gamma,\lambda]\mu\nu}_{(2)T,T}+\frac{4}{3}C_{(\gamma \ ,\mu)}^{\ \ \rho}\Gamma^{(1)T}_{\beta, \nu \lambda \rho}\Gamma^{\beta[\gamma,\lambda]\mu\nu}_{(2)T,T}
+\frac{14}{3}C_{\gamma\lambda,}^{\ \ \rho}\Gamma_{\beta,\rho\mu\nu}^{(1)T}\Gamma^{\beta[\gamma,\lambda]\mu\nu}_{(2)T,T}
\nonumber \\
+\frac{1}{2}\nabla_\gamma W_{\mu \ ,\nu}^{\ \tau \ \rho}(\Gamma^{T}_{\lambda\tau\rho}\Gamma^{[\gamma,\lambda]\mu\nu}_{(1)T}+\Gamma^{(1)[\gamma}_{T \  \ ,\lambda\tau\rho}\Gamma_{T}^{\lambda]\mu\nu})-4C_{\gamma\lambda,}^{\quad \rho}(\Gamma^{T}_{\rho\mu\nu}\Gamma^{\gamma,\lambda\mu\nu}_{(1)T}+\Gamma^{\gamma}_{{(1)}T,\rho\mu\nu}\Gamma^{\lambda\mu\nu}_{T})\nonumber\\
-3C^{\ \tau,\rho}_{\mu}(\Gamma^{T}_{\tau\rho\nu}\Gamma^{(1)\alpha\ \mu\nu}_{T \ ,\alpha}+\Gamma_{T}^{\beta \mu\nu}\Gamma_{\beta,\nu\tau\rho}^{(1)T}).\label{3.111}
\end{gather}
Then to proceed with (\ref{3.110}) we should use Weyl variations of the following two terms:
\begin{gather}
(\delta-2\sigma)B=
(\delta-2\sigma) (\nabla_\sigma \nabla_\rho W_{\alpha \mu , \beta \nu}+2K_{\sigma\rho}W_{\alpha\mu,\beta\nu}-
K_{\sigma[\alpha}W_{\mu]\rho,\beta\nu}-
K_{\sigma[\beta}W_{\nu]\rho,\alpha\mu}\nonumber \\+
g_{\rho[\alpha}K^{\tau}_{\sigma}W_{\mu]\tau,\beta\nu}+
g_{\rho[\beta}K^{\tau}_{\sigma}W_{\nu]\tau,\alpha\mu})T^{\alpha \beta \rho , \mu \nu \sigma}=
 -10\sigma_\sigma \nabla_{\rho}W_{\alpha \mu ,\beta \nu}T^{\alpha \beta \rho , \mu \nu \sigma} \nonumber \\
 +3\sigma^\tau \nabla_\tau W_{\alpha \mu,\beta \nu}T^{\alpha \beta, \mu \nu}
 -4 \sigma^\tau C_{\tau\rho , \nu} T^{\rho,\nu}
 -4\sigma_\nu C_{\mu \alpha ,\beta} T^{\alpha \beta,\mu \nu}, \label{3.112}\\
 \delta C=\delta (\Box W_{\alpha \mu ,\beta \nu}+2JW_{\alpha \mu , \beta \nu})T^{\alpha \beta,\mu \nu}=
 2 \sigma^\tau \nabla_\tau W_{\alpha \mu , \beta \nu}T^{\alpha \beta,\mu \nu}
 -16\sigma_\nu C_{\mu \alpha ,\beta} T^{\alpha \beta,\mu \nu}
 -8\sigma^\tau C_{\tau\alpha ,\nu} T^{\rho \nu}.\label{3.113}
 \end{gather}
Combining with (\ref{3.110}) we see that
\begin{gather}
4\sigma^\lambda \nabla_\mu W_{\nu \ ,\gamma}^{\ \tau \ \, \rho}T^{\gamma \mu \nu}_{\tau \rho \lambda}-
4\sigma^\gamma\nabla_{\gamma}W_{\mu \ ,\nu}^{\ \rho \ \, \tau}T^{\mu \nu}_{\rho \tau}+
24\sigma_{\mu}C^{\ \tau,\rho}_{\nu}T^{\mu \nu}_{\tau \rho}+
32\sigma^{\gamma}C_{\gamma \ ,\rho}^{\ \mu}T^{\rho}_{\mu}+\frac{2}{5}\delta(B)+
\frac{7}{5}\delta(C)\nonumber\\=8\frac{12}{5}\sigma^{\gamma}C_{\gamma \ ,\rho}^{\ \mu}T^{\rho}_{\mu}.\label{3.114}
\end{gather}
It means that after all possible cancelation we arrive to the last term in r.h.s of (\ref{3.114}).
To cancel that we need to consider similar construction for another invariant $J^{(2)}_{\mu\nu;\gamma\lambda}$ (\ref{3.94}).
Starting now from
\begin{align}
  &J^{(2)}_{\mu\nu;\gamma\lambda}K_{(T)}^{\mu\nu;\gamma\lambda}\nonumber\\
   & =  \Big[ 16\nabla_{\mu}W^{\quad\tau\quad\rho}_{\nu\quad,\gamma}
   h^{T}_{\lambda\tau\rho}
   -8C^{\quad\rho}_{(\gamma\quad,\mu)}h^{T}_{\nu\lambda\rho} +2W^{\qquad\rho}_{\gamma\lambda,\mu}\Gamma^{(1)T\,\,\,\,\tau}_{\tau;\nu\rho}\Big]
   K_{(T)}^{\mu\nu;\gamma\lambda}\nonumber\\ & +\Big[6W^{\quad(\tau\quad\rho)}_{\gamma\quad,\mu}
  \Gamma^{(1)T}_{[\nu;\tau]\lambda\rho}
+2W^{\quad\tau\quad\rho}_{\mu\quad,\nu}\Gamma^{(1)T}_{[\gamma;\lambda]\tau\rho}
\Big]K_{(T)}^{\mu\nu;\gamma\lambda}\label{3.115}
\end{align}
instead of (\ref{3.98}) we have
\begin{gather}
-\frac{1}{6}W_{\gamma\lambda,\mu}^{\ \quad \rho}\tilde{\gamma}_{\beta,\nu\rho}\Gamma^{\beta[\gamma,\lambda]\mu\nu}_{(2)T,T}+
\frac{4}{3}W_{\gamma \lambda,\mu}^{\quad \ \rho}t^{T}_{\rho \nu \beta}\Gamma^{\beta[\gamma,\lambda]\mu\nu}_{(2)T,T}-
W_{\gamma \lambda,\beta}^{\quad \ \rho}t^{T}_{\rho \mu \nu}\Gamma^{\beta[\gamma,\lambda]\mu \nu}_{(2)T,T}\nonumber \\+
W_{\mu\ ,\nu}^{\ \tau \ \, \rho}t^{T}_{\tau \rho \beta}(\frac{10}{3}\tilde{\gamma}^{\lambda,\mu\nu}-\frac{4}{3}\tilde{\gamma}^{\mu,\nu\lambda})+
\frac{2}{3}C_{(\gamma \ , \nu)}^{\ \ \rho}\Gamma^{(1)T}_{\beta,\lambda\mu\rho}\delta\Gamma^{\beta[\gamma,\lambda]\mu\nu}_{(2)T;T}-
\frac{4}{3}\nabla_{\mu}W_{\nu \ \,,\gamma}^{\ \tau \ \, \rho}\Gamma^{(1)T}_{\beta,\tau \rho \lambda}\delta \Gamma^{\beta[\gamma,\lambda]\mu\nu}_{(2)T,T}.\label{3.116}
\end{gather}
Then using again (\ref{3.100}) with $A=\frac{1}{2}$ we obtain instead of (\ref{3.101})
\begin{gather}
\frac{2}{3}C_{(\gamma \ , \nu)}^{\ \ \rho}\Gamma^{(1)T}_{\beta,\lambda\mu\rho}\delta\Gamma^{\beta[\gamma,\lambda]\mu\nu}_{(2)T;T}-
\frac{4}{3}\nabla_{\mu}W_{\nu \ \,,\gamma}^{\ \tau \ \, \rho}\Gamma^{(1)T}_{\beta,\tau \rho \lambda}\delta \Gamma^{\beta[\gamma,\lambda]\mu\nu}_{(2)T,T}+
\frac{1}{3}C^{\quad \rho}_{\gamma \lambda}\Gamma_{\beta,\rho \mu \nu}^{(1)T}\delta\Gamma^{\beta[\gamma,\lambda]\mu\nu}_{(2)T,T}\nonumber \\-
\frac{4}{3}W_{\gamma \lambda,\beta}^{\quad \ \rho}t^{T}_{\rho \mu \nu}\Gamma^{\beta[\gamma,\lambda]\mu \nu}_{(2)T,T}+
\frac{4}{3}W_{\gamma \lambda,\mu}^{\quad \ \rho}t^{T}_{\rho \nu \beta}\Gamma^{\beta[\gamma,\lambda]\mu\nu}_{(2)T,T}+
W_{\mu\ ,\nu}^{\ \tau \ \, \rho}t^{T}_{\tau \rho \beta}(\frac{10}{3}\Gamma^{(2) \beta; \alpha \mu \nu}_{T\, \alpha}-\frac{4}{3}\Gamma^{(2) \mu; \alpha \nu \beta}_{T \, \alpha}).\label{3.117}
\end{gather}
Then the same miraculous cancelation leads instead of (\ref{3.110}) to the following reminder:
\begin{gather}
3\sigma^\lambda \nabla_\mu W_{\nu \ ,\gamma}^{\ \tau \ \, \rho}T^{\gamma \mu \nu}_{\tau \rho \lambda}-
2\sigma^\gamma\nabla_{\gamma}W_{\mu \ ,\nu}^{\ \rho \ \, \tau}T^{\mu \nu}_{\rho \tau}+
10\sigma_{\mu}C^{\ \tau,\rho}_{\nu}T^{\mu \nu}_{\tau \rho}+
8\sigma^{\gamma}C_{\gamma \ ,\rho}^{\ \mu}T^{\rho}_{\mu},\label{3.118}
\end{gather}
and corresponding integrated terms instead of (\ref{3.111}) is
\begin{gather}
-L^{(2)}=
+ \frac{1}{9} W_{\mu\ ,\nu}^{\ \tau \ \, \rho}\Gamma^{(2)T;T}_{\beta[\gamma,\lambda]\tau \rho} \Gamma_{(2)T;T}^{\beta[\gamma,\lambda]\mu \nu}
+ \frac{1}{3}W_{\gamma\ ,\mu}^{\, ( \tau \ \, \rho)}\Gamma^{(2)T;T}_{\beta[\nu,\tau]\lambda \rho} \Gamma_{(2)T;T}^{\beta[\gamma,\lambda]\mu \nu}
+ \frac{1}{36}W_{\gamma \lambda ,}^{\ \ \, \tau \rho }\Gamma^{(2)T;T}_{\beta[\tau,\rho]\mu \nu} \Gamma_{(2)T;T}^{\beta[\gamma,\lambda]\mu \nu}\nonumber\\- \frac{1}{96}W_{\mu\ ,\nu}^{\ \tau \ \, \rho}\Gamma^{T}_{\tau \rho \lambda}\Gamma_{T}^{\mu \nu \lambda}\nonumber\\+ \frac{1}{6}W_{\gamma \lambda , \mu}^{\ \ \ \ \rho}\Gamma^{T}_{\rho \nu \beta}\Gamma^{\beta [\gamma , \lambda] \mu \nu}_{(1)T}- \frac{1}{6}W_{\gamma \lambda , \beta}^{\ \ \ \ \rho}\Gamma^{T}_{\rho \mu \nu}\Gamma^{\beta [\gamma , \lambda] \mu \nu}_{(1)T}
+ \frac{1}{8} W_{\mu\ ,\nu}^{\ \tau \ \, \rho}\Gamma^{T}_{\tau \rho \beta}(\frac{10}{3}\Gamma^{(2) \beta; \alpha \mu \nu}_{T \, \alpha}-\frac{4}{3}\Gamma^{(2) \mu; \alpha \nu \beta}_{T\, \alpha})\nonumber\\+\frac{4}{3}\nabla_{\mu} W_{\nu\ ,\gamma}^{\ \tau \ \, \rho}\Gamma^{(1)T}_{\beta,\lambda \tau \rho}\Gamma^{\beta[\gamma,\lambda]\mu\nu]}_{(2)T;T}
-\frac{2}{3}C_{(\gamma, \ \nu)}^{\ \ \rho}\Gamma^{(1)T}_{\beta,\lambda \mu \rho} \Gamma_{(2)T;T}^{\beta[\gamma,\lambda]\mu \nu}
- \frac{1}{3}C_{\gamma \lambda ,}^{\ \ \, \rho}\Gamma^{(1)T}_{\beta,\rho \mu \nu}\Gamma_{(2)T;T}^{\beta[\gamma,\lambda]\mu \nu} \nonumber \\
- \frac{1}{2}\nabla_{\mu}W_{\nu\ ,\gamma}^{\ \tau \ \, \rho}(\Gamma^{T}_{\tau \rho \lambda}\Gamma^{[\gamma,\lambda]\mu \nu}_{(1)T}+\Gamma^{(1)T[\gamma}_{\ \ \ \ \ \,\tau \rho \lambda}\Gamma_{T}^{\lambda]\mu \nu})
- \frac{1}{8}\nabla^{\mu}W_{\nu\ ,\gamma}^{\ \tau \ \, \rho}(\Gamma^{T}_{\tau \rho \mu}\Gamma^{(1)\alpha,  \ \gamma \nu }_{T  \ \ \ \alpha}+\Gamma^{(1)T}_{\beta,\tau \rho \mu}\Gamma_{T}^{\beta \gamma \nu})\nonumber \\
+\frac{1}{2}C_{\gamma \lambda,}^{\ \ \ \rho}(\Gamma^{(1)T;\gamma}_{\ \ \ \ \ ,\rho \mu \nu}\Gamma_{T}^{\lambda \mu \rho}+\Gamma^{\gamma ,\lambda \mu \nu}_{(1)T}\Gamma^{T}_{\rho \mu \nu})
+ \frac{3}{4}C^{\ \  \rho,\tau}_{\mu}(\Gamma^{T}_{\rho \tau \nu}\Gamma^{(1)T\alpha \mu\nu}_{ \ \ \alpha,}+\Gamma_{T}^{\mu\nu \beta}\Gamma^{(1)T}_{\beta, \rho \tau \nu}).\nonumber \\
\label{3.119}
\end{gather}
Then in the similar way (see (\ref{3.115})) we can write integration rules for reminder (\ref{3.119})
\begin{gather}
3\sigma^\lambda \nabla_\mu W_{\nu \ ,\gamma}^{\ \tau \ \, \rho}T^{\gamma \mu \nu}_{\tau \rho \lambda}-
2\sigma^\gamma\nabla_{\gamma}W_{\mu \ ,\nu}^{\ \rho \ \, \tau}T^{\mu \nu}_{\rho \tau}+
10\sigma_{\mu}C^{\ \tau,\rho}_{\nu}T^{\mu \nu}_{\tau \rho}+
8\sigma^{\gamma}C_{\gamma \ ,\rho}^{\ \mu}T^{\rho}_{\mu}+\frac{3}{10}\delta(B)+
\frac{11}{20}\delta(C)\nonumber \\=\frac{12}{5}\sigma^{\gamma}C_{\gamma \ ,\rho}^{\ \mu}T^{\rho}_{\mu}.\label{3.120}
\end{gather}
Comparing last equation with the (\ref{3.115}) we see that the following expression will be Weyl Invariant
\begin{gather}
I_{W}=\big[J^{(1)}_{\mu\nu;\gamma\lambda}
-8J^{(2)}_{\mu\nu;\gamma\lambda}
+I^{(4)}_{\mu\nu;\gamma\lambda}\big]K_{(T)}^{\mu\nu;\gamma\lambda}
+L^{(1)}-8L^{(2)}-2B-3C .\label{3.121}
\end{gather}
Then multiplying (\ref{3.119}) by $8$ and summing with (\ref{3.11}) we get
\begin{gather}
L^{(1)}-8L^{(2)}=\frac{2}{3}W_{\mu \ ,\nu}^{\ \tau \ \, \rho}\Gamma^{(2)T,T}_{\beta[\gamma,\lambda]\tau \rho}\Gamma^{\beta[\gamma,\lambda]\mu\nu}_{(2)T,T}
+\frac{22}{9}W_{\gamma \ ,\mu}^{\ (\tau \ \, \rho)}\Gamma^{(2)T,T}_{\beta[\nu,\tau]\lambda \rho}\Gamma^{\beta[\gamma,\lambda]\mu\nu}_{(2)T,T}
-\frac{1}{6}W_{\gamma\lambda,}^{\ \ \tau \rho}\Gamma^{(2)T,T}_{\beta[\tau,\rho]\mu\nu}\Gamma^{\beta[\gamma,\lambda]\mu\nu}_{(2)T,T}\nonumber \\
-\frac{1}{8}W_{\mu \ ,\nu}^{\ \tau \ \, \rho}\Gamma^{T}_{\tau \rho \lambda}\Gamma_{T}^{\mu \nu \lambda}\nonumber\\
+W_{\gamma\lambda,\mu}^{\ \ \ \rho}\Gamma^{T}_{\rho\nu\beta}\Gamma^{\beta[\gamma,\lambda]\mu\nu}_{(2)T,T}
-\frac{2}{3}W_{\gamma\lambda,\beta}^{\ \ \ \ \rho}\Gamma^{T}_{\rho\mu\nu}\Gamma^{\beta[\gamma,\lambda]\mu\nu}_{(2)T,T}
+\frac{17}{6}W_{\mu \ \ ,\nu}^{\ \tau \ \ \rho}\Gamma^{T}_{\tau\rho\beta}\Gamma^{(2)\ \beta,\mu\nu\alpha}_{T\alpha}
-\frac{4}{3}W_{\mu\ ,\nu}^{\ \tau \ \, \rho}\Gamma^{T}_{\tau \rho \beta}\Gamma^{(2) \mu; \alpha \nu \beta}_{T\, \alpha} \nonumber\\-\left(\frac{4}{3}\nabla_{\gamma}W_{\mu \ ,\nu}^{\ \tau \ \, \rho}-\frac{32}{3}\nabla_{\mu} W_{\nu\ ,\gamma}^{\ \tau \ \, \rho}\right)\Gamma^{(1)T}_{\beta,  \lambda \tau \rho}\Gamma^{\beta[\gamma,\lambda]\mu\nu}_{(2)T,T}
-4C_{(\gamma \ ,\mu)}^{\ \ \rho}\Gamma^{(1)T}_{\beta, \nu \lambda \rho}\Gamma^{\beta[\gamma,\lambda]\mu\nu}_{(2)T,T}
+2C_{\gamma\lambda,}^{\ \ \rho}\Gamma_{\beta,\rho\mu\nu}^{(1)T}\Gamma^{\beta[\gamma,\lambda]\mu\nu}_{(2)T,T}\nonumber\\+\left(\frac{1}{2}\nabla_{\gamma} W_{\mu \ ,\nu}^{\ \tau \ \rho}-4\nabla_{\mu}W_{\nu\ ,\gamma}^{\ \tau \ \, \rho}\right)(\Gamma^{T}_{\lambda\tau\rho}\Gamma^{[\gamma,\lambda]\mu\nu}_{(1)T}+\Gamma^{(1)[\gamma}_{T \  \ ,\lambda\tau\rho}\Gamma_{T}^{\lambda]\mu\nu})-\nabla^{\mu}W_{\nu\ ,\gamma}^{\ \tau \ \, \rho}(\Gamma^{T}_{\tau \rho \mu}\Gamma^{(1)\alpha,  \ \gamma \nu }_{T  \ \ \ \alpha}+\Gamma^{(1)T}_{\beta,\tau \rho \mu}\Gamma_{T}^{\beta \gamma \nu}) \nonumber \\
+3C^{\ \tau,\rho}_{\mu}(\Gamma^{T}_{\tau\rho\nu}\Gamma^{(1)\alpha\ \mu\nu}_{T \ ,\alpha}+\Gamma_{T}^{\beta \mu\nu}\Gamma_{\beta,\nu\tau\rho}^{(1)T}). \label{3.122}
\end{gather}
In the same way we can collect our starting terms (\ref{3.96}), (\ref{3.115}) and (\ref{3.101}) and obtain
\begin{align}
&\big[J^{(1)}_{\mu\nu;\gamma\lambda}
-8J^{(2)}_{\mu\nu;\gamma\lambda}
+I^{(4)}_{\mu\nu;\gamma\lambda}\big]K_{(T)}^{\mu\nu;\gamma\lambda}\nonumber\\
& = \Big[-12 W^{\quad\tau\quad\rho}_{\mu\quad,\nu}\Gamma^{(1)T}_{[\gamma;\lambda]\tau\rho}-44W^{\quad(\tau\quad\rho)}_{\gamma\quad,\mu}
\Gamma^{(1)T}_{[\nu;\tau]\lambda\rho} +3W^{\quad\tau\rho}_{\gamma\lambda,}\Gamma^{(1)T}_{[\tau;\rho]\mu\nu}-20W^{\qquad\rho}_{\gamma\lambda,\mu}\Gamma^{(1)T\,\,\,\,\tau}_{\tau;\nu\rho}\Big]K_{(T)}^{\mu\nu;\gamma\lambda}\nonumber\\
&+16\Big[\nabla_{\gamma}W^{\quad\tau\quad\rho}_{\mu\quad,\nu}
h^{T}_{\lambda\tau\rho}-8\nabla_{\mu}W^{\quad\tau\quad\rho}_{\nu\quad,\gamma}
h^{T}_{\lambda\tau\rho}
+3C^{\quad\rho}_{(\gamma\quad,\mu)}h^{T}_{\nu\lambda\rho}-\frac{3}{2}C^{\quad\rho}_{\gamma\lambda,}h^{T}_{\rho\mu\nu}\Big]
K_{(T)}^{\mu\nu;\gamma\lambda}.\label{3.123}
\end{align}
Then we can do the following simplifications:
First of all note that the following modified  by gauge invariant in zero order on background curvature Christoffel symbol
\begin{equation}\label{3.124}
\tilde{\Gamma}^{\beta[\gamma,\lambda]\mu\nu}_{(2)T,T}=
\Gamma^{\beta[\gamma,\lambda]\mu\nu}_{(2)T,T}-\frac{3}{8}\left( g^{\beta[\gamma}\Gamma_{T}^{\lambda]\mu\nu}-\frac{1}{4}g^{[\gamma(\mu}\Gamma_{T}^{\nu)\lambda]\beta}\right)
\end{equation}
transforms without third line in (\ref{3.70}) but with the same in zero order on curvature gauge variation. Shifting all $\Gamma^{\beta[\gamma,\lambda]\mu\nu}_{(2)T,T}$ to the $\tilde{\Gamma}^{\beta[\gamma,\lambda]\mu\nu}_{(2)T,T}$ in (\ref{3.122}) we cancel second and third line there and modify other terms bilinear on first and second Christoffel symbols. Then last modification could be done after the change to the completely traceless object
\begin{equation}\label{3.125}
\Gamma_{(1)T;T}^{[\gamma;\lambda]\mu\nu}= \Gamma_{;T}^{(1)[\gamma;\lambda]\mu\nu}
- \frac{1}{4}g^{[\gamma(\mu}\Gamma^{(1)\alpha,  \ \nu)\lambda]) }_{T  \ \ \ \alpha}
\end{equation}
in the first term of fifth line of (\ref{3.122}). After this transformations we obtain:
\begin{gather}
L^{(1)}-8L^{(2)}=\frac{2}{3}W_{\mu \ ,\nu}^{\ \tau \ \, \rho}\tilde{\Gamma}^{(2)T,T}_{\beta[\gamma,\lambda]\tau \rho}\tilde{\Gamma}^{\beta[\gamma,\lambda]\mu\nu}_{(2)T,T}
+\frac{22}{9}W_{\gamma \ ,\mu}^{\ (\tau \ \, \rho)}\tilde{\Gamma}^{(2)T,T}_{\beta[\nu,\tau]\lambda \rho}\tilde{\Gamma}^{\beta[\gamma,\lambda]\mu\nu}_{(2)T,T}
-\frac{1}{6}W_{\gamma\lambda,}^{\ \ \tau \rho}\tilde{\Gamma}^{(2)T,T}_{\beta[\tau,\rho]\mu\nu}\tilde{\Gamma}^{\beta[\gamma,\lambda]\mu\nu}_{(2)T,T}\nonumber \\
-\left(\frac{4}{3}\nabla_{\gamma}W_{\mu \ ,\nu}^{\ \tau \ \, \rho}-\frac{32}{3}\nabla_{\mu} W_{\nu\ ,\gamma}^{\ \tau \ \, \rho}\right)\Gamma^{(1)T}_{\beta,  \lambda \tau \rho}\tilde{\Gamma}^{\beta[\gamma,\lambda]\mu\nu}_{(2)T,T}
-4C_{(\gamma \ ,\mu)}^{\ \ \rho}\Gamma^{(1)T}_{\beta, \nu \lambda \rho}\tilde{\Gamma}^{\beta[\gamma,\lambda]\mu\nu}_{(2)T,T}
+2C_{\gamma\lambda,}^{\ \ \rho}\Gamma_{\beta,\rho\mu\nu}^{(1)T}\tilde{\Gamma}^{\beta[\gamma,\lambda]\mu\nu}_{(2)T,T}\nonumber\\+\left(\frac{1}{2}\nabla_{\gamma} W_{\mu \ ,\nu}^{\ \tau \ \rho}-4\nabla_{\mu}W_{\nu\ ,\gamma}^{\ \tau \ \, \rho}\right)\Gamma^{T}_{\lambda\tau\rho}\Gamma^{[\gamma,\lambda]\mu\nu}_{(1)T;T}
+3C^{\ \rho}_{\mu\ \ ,\gamma}\Gamma^{T}_{\rho\lambda\nu}\Gamma^{[\gamma;\lambda]\mu\nu}_{(1)T;T} .\label{3.126}
\end{gather}
Then using the following relation:
\begin{align}
4C_{(\gamma \ ,\mu)}^{\ \ \rho}\Gamma^{(1)T}_{\beta, \nu \lambda \rho}\tilde{\Gamma}^{\beta[\gamma,\lambda]\mu\nu}_{(2)T,T}
-2C_{\gamma\lambda,}^{\ \ \rho}\Gamma_{\beta,\rho\mu\nu}^{(1)T}\tilde{\Gamma}^{\beta[\gamma,\lambda]\mu\nu}_{(2)T,T}=8 C^{\ \rho}_{\mu\ \ ,\gamma}\Gamma^{(1)T}_{\beta, \nu \lambda \rho}\tilde{\Gamma}^{\beta[\gamma,\lambda]\mu\nu}_{(2)T,T} \label{3.127}
\end{align}
we arrive to the following compact two line formula:
\begin{gather}
L^{(1)}-8L^{(2)}=\frac{2}{3}W_{\mu \ ,\nu}^{\ \tau \ \, \rho}\tilde{\Gamma}^{(2)T,T}_{\beta[\gamma,\lambda]\tau \rho}\tilde{\Gamma}^{\beta[\gamma,\lambda]\mu\nu}_{(2)T,T}
+\frac{22}{9}W_{\gamma \ ,\mu}^{\ (\tau \ \, \rho)}\tilde{\Gamma}^{(2)T,T}_{\beta[\nu,\tau]\lambda \rho}\tilde{\Gamma}^{\beta[\gamma,\lambda]\mu\nu}_{(2)T,T}
-\frac{1}{6}W_{\gamma\lambda,}^{\ \ \tau \rho}\tilde{\Gamma}^{(2)T,T}_{\beta[\tau,\rho]\mu\nu}\tilde{\Gamma}^{\beta[\gamma,\lambda]\mu\nu}_{(2)T,T}\nonumber \\
-\Big[\nabla_{\gamma}W_{\mu \ ,\nu}^{\ \tau \ \, \rho}-8\nabla_{\mu} W_{\nu\ ,\gamma}^{\ \tau \ \, \rho}+6C^{\ \rho}_{\mu\ \ ,\gamma}\delta^{\tau}_{\nu}\Big]\left( \frac{4}{3}\Gamma^{(1)T}_{\beta,  \lambda \tau \rho}\tilde{\Gamma}^{\beta[\gamma,\lambda]\mu\nu}_{(2)T,T}
-\frac{1}{2}\Gamma^{T}_{\lambda\tau\rho}\Gamma^{[\gamma,\lambda]\mu\nu}_{(1)T;T}\right).
\label{3.128}
\end{gather}
Then applying (\ref{3.125}) and (\ref{3.127}) to the first and second line of  (\ref{3.123}) we obtain corresponding cancelation of the last trace term of first line and reduction of terms in second line with the same combination of derivatives of Weyl tensors in brackets:
\begin{align}
&\big[J^{(1)}_{\mu\nu;\gamma\lambda}
-8J^{(2)}_{\mu\nu;\gamma\lambda}
+I^{(4)}_{\mu\nu;\gamma\lambda}\big]K_{(T)}^{\mu\nu;\gamma\lambda}\nonumber\\
& = \Big[-12 W^{\quad\tau\quad\rho}_{\mu\quad,\nu}\Gamma^{(1)T,T}_{[\gamma;\lambda]\tau\rho}-44W^{\quad(\tau\quad\rho)}_{\gamma\quad,\mu}
\Gamma^{(1)T,T}_{[\nu;\tau]\lambda\rho} +3W^{\quad\tau\rho}_{\gamma\lambda,}\Gamma^{(1)T,T}_{[\tau;\rho]\mu\nu}\Big]K_{(T)}^{\mu\nu;\gamma\lambda}\nonumber\\
&+16\Big[\nabla_{\gamma}W^{\quad\tau\quad\rho}_{\mu\quad,\nu}
-8\nabla_{\mu}W^{\quad\tau\quad\rho}_{\nu\quad,\gamma}
+6C^{\quad\rho}_{\mu\quad,\gamma}\delta^{\tau}_{\nu}\Big]h^{T}_{\lambda\tau\rho}
K_{(T)}^{\mu\nu;\gamma\lambda}.\label{3.129}
\end{align}
Combining (\ref{3.129}) and (\ref{3.128}) we arrive to the result that the following final expression
\begin{align}
L^{W\Gamma\Gamma}_{-4}=
&\Big[(J^{(1)}_{\mu\nu;\gamma\lambda}
-8J^{(2)}_{\mu\nu;\gamma\lambda}
+I^{(4)}_{\mu\nu;\gamma\lambda})K_{(T)}^{\mu\nu;\gamma\lambda}+L^{(1)}-8L^{(2)}-2B-3C\Big]\nonumber\\
&=\frac{2}{3}W_{\mu \ ,\nu}^{\ \tau \ \, \rho}\tilde{\Gamma}^{(2)T,T}_{\beta[\gamma,\lambda]\tau \rho}\tilde{\Gamma}^{\beta[\gamma,\lambda]\mu\nu}_{(2)T,T}
+\frac{22}{9}W_{\gamma \ ,\mu}^{\ (\tau \ \, \rho)}\tilde{\Gamma}^{(2)T,T}_{\beta[\nu,\tau]\lambda \rho}\tilde{\Gamma}^{\beta[\gamma,\lambda]\mu\nu}_{(2)T,T}
-\frac{1}{6}W_{\gamma\lambda,}^{\ \ \tau \rho}\tilde{\Gamma}^{(2)T,T}_{\beta[\tau,\rho]\mu\nu}\tilde{\Gamma}^{\beta[\gamma,\lambda]\mu\nu}_{(2)T,T}\nonumber \\
&-\Big[\nabla_{\gamma}W_{\mu \ ,\nu}^{\ \tau \ \, \rho}-8\nabla_{\mu} W_{\nu\ ,\gamma}^{\ \tau \ \, \rho}+6C^{\ \rho}_{\mu\ \ ,\gamma}\delta^{\tau}_{\nu}\Big]\left( \frac{4}{3}\Gamma^{(1)T}_{\beta,  \lambda \tau \rho}\tilde{\Gamma}^{\beta[\gamma,\lambda]\mu\nu}_{(2)T,T}
-\frac{1}{2}\Gamma^{T}_{\lambda\tau\rho}\Gamma^{[\gamma,\lambda]\mu\nu}_{(1)T;T}-16h^{T}_{\lambda\tau\rho}K_{(T)}^{\mu\nu;\gamma\lambda}\right)\nonumber\\
& -\Big[12 W^{\quad\tau\quad\rho}_{\mu\quad,\nu}\Gamma^{(1)T,T}_{[\gamma;\lambda]\tau\rho}+44W^{\quad(\tau\quad\rho)}_{\gamma\quad,\mu}
\Gamma^{(1)T,T}_{[\nu;\tau]\lambda\rho} -3W^{\quad\tau\rho}_{\gamma\lambda,}\Gamma^{(1)T,T}_{[\tau;\rho]\mu\nu}\Big]
K_{(T)}^{\mu\nu;\gamma\lambda}\nonumber\\
& -2\Big[(\nabla^\sigma \nabla_\rho+4K^{\sigma}_{\rho})W^{\,\,\,\mu\,\,\,\,\,\nu}_{\alpha\,\, , \beta}\Big]T^{\alpha \beta \rho}_{\mu \nu \sigma} + \Big[4K^{\mu\tau}W_{\alpha\tau,\beta}^{\quad\,\,\nu}
 -3(\Box+2J)W_{\alpha\,\,\,, \beta}^{\,\,\,\mu\,\,\,\,\nu})\Big]T^{\alpha \beta}_{\mu \nu}\label{3.131}
\end{align}
is weight -4 primary field and can be used as a Weyl invariant Lagrangian.

\begin{footnotesize}

\end{footnotesize}

\end{document}